\documentclass[aip,jcp,reprint,amsmath,amssymb,nofootinbib]{revtex4-2}

\usepackage{graphicx}
\usepackage{dcolumn}
\usepackage{bm}
\usepackage{xcolor}
\usepackage[colorlinks=true,linkcolor=blue,citecolor=blue,urlcolor=blue]{hyperref}

\graphicspath{{figures/}}

\newcommand{\avg}[1]{\left\langle #1 \right\rangle}
\newcommand{\kBT}{k_\mathrm{B}T}
\newcommand{\diff}{\mathrm{d}}

\newcommand{\dmat}[1]{\underline{\underline{\mathbf{#1}}}}

\makeatletter
\newcounter{sipart}\renewcommand{\thesipart}{\Alph{sipart}}
\newcommand{\sipart}[1]{\par\vspace{2.5\baselineskip}\refstepcounter{sipart}%
  \noindent{\Large\bfseries Part \thesipart:\quad #1}\par\nopagebreak\vspace{0.5\baselineskip}%
  \@afterindentfalse\@afterheading}
\newcounter{sinote}[sipart]\renewcommand{\thesinote}{\thesipart\arabic{sinote}}
\newcommand{\sinote}[1]{\par\vspace{2\baselineskip}\refstepcounter{sinote}%
  \noindent{\large\bfseries\thesinote\quad #1}\par\nopagebreak\vspace{0.5\baselineskip}%
  \@afterindentfalse\@afterheading}
\newcounter{sisub}[sinote]\renewcommand{\thesisub}{\thesinote.\arabic{sisub}}
\newcommand{\sisub}[1]{\par\vspace{1.25\baselineskip}\refstepcounter{sisub}%
  \noindent{\bfseries\thesisub\quad #1}\par\nopagebreak\vspace{0.25\baselineskip}%
  \@afterindentfalse\@afterheading}
\makeatother

\newcommand{\hn}[1]{\hat n_{#1}}
\newcommand{\Ihc}{\mathcal I_{hc}}
\newcommand{\Ps}{\bm\Psi}
\newcommand{\A}{\underline{\underline{\mathbf A}}}
\newcommand{\F}{\underline{\underline{\mathbf F}}}
\newcommand{\K}{\underline{\underline{\mathbf K}}}

\newcommand{\dd}{\,du}
\newcommand{\hR}{\hat{\mathbf R}}
\newcommand{\hf}{\hat{\mathbf f}}

\begin{document}

\title{Analytical Rouse theory reveals how a loop shapes the structure and dynamics of a flexible polymer}

\author{Rishan Gobse}
\altaffiliation{These authors contributed equally to this work.}
\affiliation{Department of Chemical Engineering, Indian Institute of Technology Bombay, Mumbai 400076, India}

\author{Rishit Kesharwani}
\altaffiliation{These authors contributed equally to this work.}
\affiliation{Department of Chemical Engineering, Indian Institute of Technology Bombay, Mumbai 400076, India}

\author{Sayantan Dutta\textsuperscript{\normalfont\textcolor{blue}{a},}}
\email[Author to whom correspondence should be addressed: ]{\mbox{sayantan.dutta@iitb.ac.in}}
\affiliation{Department of Chemical Engineering, Indian Institute of Technology Bombay, Mumbai 400076, India}

\begin{abstract}
Architectural loops, such as those formed by cohesin in chromatin, constrain the conformations and dynamics of biopolymers. The Rouse model of a flexible polymer is widely used to describe such dynamics, but exact solutions are available only for symmetric architectures and not for a loop at an arbitrary position along a chain. Here, we develop an analytical Rouse theory for a flexible polymer with a single loop of arbitrary size and position. We represent the looped chain as a star polymer with arms of unequal lengths and introduce a modal decomposition that depends only on the positions of the two loop anchors. We derive a closed-form expression for the equilibrium statistics of the modes and an exact expression for their dynamics in terms of the loop topology alone, which ultimately yield the structural and dynamical properties of the chain without simulation. A loop compacts the chain and enhances long-range contacts. In the motion of a tagged locus, a loop leaves a transient slowdown, a dip of the local exponent of the mean squared displacement below the Rouse value, whose onset and minimum are set by the distance of the locus from the nearest anchor and by the size of the loop. Even at short times, where all loci move with the same exponent, the ratios of the mean squared displacements of a few loci placed around an anchor are predictive of the loop architecture. Altogether, this framework provides a basis for interpreting the motion of tagged loci in terms of the loops that constrain them.
\end{abstract}

\maketitle

\section{Introduction}
\label{sec:introduction}
\vspace{-10 pt}
Many biopolymers are not simple linear chains: constraints that hold distant segments together give them an architecture that affects both their conformations and their dynamics. More generally, constraints on a flexible chain, such as entropic barriers along its path or a pore it must thread, control how it moves~\cite{muthukumar1989,muthukumar1999}. Chromatin is one example. During interphase, protein complexes such as cohesin bring pairs of genomic loci into contact and fold the chromatin fiber into loops~\cite{ganji2018,davidson2019,kim2019,sanborn2015,fudenberg2016,mirny2019}, which modulate the communication between distal genomic elements and are implicated in gene regulation~\cite{dekker2016}. These loops have been studied with chromosome conformation capture techniques such as Hi-C~\cite{liebermanaiden2009,dixon2012,nora2012,rao2014}, chromatin tracing~\cite{bintu2018,mateo2019}, and live-cell imaging of tagged loci~\cite{weber2010,hajjoul2013,gabriele2022,mach2022,bruckner2023}, as well as with polymer simulations~\cite{barbieri2012,sanborn2015,fudenberg2016,dipierro2016,brackley2017,nuebler2018} and theoretical models~\cite{mateoslangerak2009,amitai2017,shukron2017,socol2019,polovnikov2023,yuan2024}. Imaging studies indicate that a given loop is present in only a fraction of cells at any time~\cite{gabriele2022,mach2022}, so ensemble-averaged contact maps do not show where a loop is in an individual cell. 

The flexible-polymer, or Rouse, model~\cite{rouse1953,doi1988} is widely used to describe such dynamics. It treats the polymer as a Gaussian chain of beads connected by harmonic springs in a viscous medium and neglects hydrodynamic and excluded-volume interactions, yet it captures the subdiffusive motion of chromosomal loci~\cite{weber2010,hajjoul2013} and underlies many models of chromatin dynamics~\cite{shukron2017,socol2019,dutta2023,dutta2024,yuan2024}. Its tractability comes from its normal modes, which for a linear chain decouple the equations of motion into independent Ornstein--Uhlenbeck processes~\cite{doi1988}. We have previously used this framework to reconstruct chromatin connectivity from live images of tagged loci~\cite{dutta2023} and to describe the effect of local active fluctuations~\cite{ghosh2022,dutta2024}, and extended it to chains with heterogeneous extensibility, for which the modes are correlated~\cite{saini2025}.

Branched architectures have been studied since the work of Ham~\cite{ham1957} and of Zimm and Kilb~\cite{zimm1959} on star polymers, the Rouse dynamics of fractal connectivity has been analysed~\cite{muthukumar1985}, and generalized Gaussian structures treat arbitrary topologies numerically~\cite{gurtovenko2005}. Exact dynamical eigenmodes have been derived for star and tadpole polymers, the latter describing a loop at one end of a chain~\cite{keesman2013}. These exact results, however, assume arms of equal length and a junction friction proportional to the number of arms~\cite{keesman2013,uneyama2024}. A loop at an arbitrary position divides the chain into a loop and two tails of generally unequal lengths, for which the symmetric-star modes no longer diagonalize the dynamics. The effects of such loops have therefore been studied mainly through simulations~\cite{polovnikov2023} and through bead--spring models solved numerically~\cite{yuan2024}. These approaches can treat many loops and dynamic loop extrusion, but the effect of a loop follows only numerically.

In this work, we develop an analytical Rouse theory for a chain with a single loop at an arbitrary position. We represent the looped chain as a four-arm star and introduce a modal decomposition that accounts for unequal arm lengths exactly, at the cost of orthogonality. The modes are coupled through a linear non-symmetric equation. We obtain the equilibrium mode--mode covariance in closed form and the time-dependent covariance from the drift matrix. These correlations determine the structural and dynamical descriptors of the chain as functions of the two anchor positions alone.  Finally, we test whether the short-time motion of a few loci is predictive of the loop architecture.

\section{Mathematical framework of a flexible polymer with arbitrary loop position}
\label{sec:model}

\begin{figure}[b]
  \centering
  \includegraphics{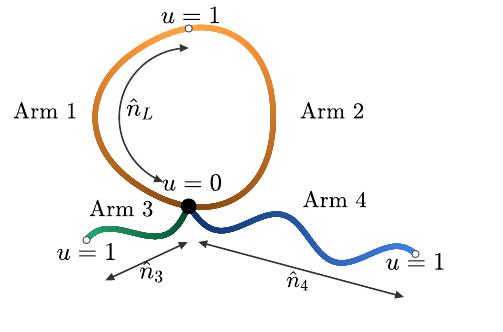}
  \caption{Schematic of a Rouse chain with a single loop, represented as four arms joined at a junction (black circle) located at the loop anchor. Arms 1 and 2 (orange) form the loop and are joined at the loop centre; each has length $\hat n_L$. Arms 3 (green) and 4 (blue) are the two free tails, of lengths $\hat n_3$ and $\hat n_4$. Lengths are in units of the total contour length $N_{tot}$. On each arm, the rescaled contour coordinate $u$ runs from $0$ at the junction to $1$ at the loop centre or at the free end.}
  \label{fig:schematic}
\end{figure}

We describe chromatin as a flexible polymer in a viscous medium using the Rouse model~\cite{rouse1953,doi1988}. A loop, such as one held by cohesin at CTCF-bound anchors~\cite{gabriele2022}, brings two loci into contact. We represent the looped chain as four Rouse arms joined at a massless junction located at the loop anchor: two loop arms of equal length, closed onto each other at their far ends, and two free tails (Fig.~\ref{fig:schematic}). We label each locus by its contour fraction $s\in[0,1]$, its position along the unlooped chain of total length $N_{tot}$ measured from one end ($s=0$) to the other ($s=1$). If the anchors sit at $s_a< s_b$, the arm lengths in units of $N_{tot}$ are
\begin{equation}
\hat{n}_3 = s_a, \qquad
\hat{n}_L = \frac{s_b-s_a}{2}, \qquad
\hat{n}_4 = 1-s_b,
\label{eq:fractions}
\end{equation}
for the first tail, each loop arm, and the second tail, respectively. The two anchor fractions therefore fix the topology of the chain completely.

\subsection{Equations of motion}
\label{sec:eom}

Each arm $a=1,\dots,4$ is a continuous Gaussian thread in three dimensions, with position $\mathbf{R}_a(n,t)\in\mathbb{R}^3$ and the contour coordinate $n\in[0,N_a]$ measured from the junction. With Kuhn length $b$ and viscous drag coefficient $\zeta$ per unit contour length, each arm obeys the Langevin equation
\begin{equation}
\zeta\,\partial_t\mathbf{R}_a = \frac{3\kBT}{b^2}\,\partial_n^2\mathbf{R}_a + \mathbf{f}_a,
\label{eq:langevin}
\end{equation}
where the thermal noise is Gaussian, independent in the three Cartesian directions, and obeys the fluctuation--dissipation theorem,
\begin{equation}
\avg{\mathbf{f}_a(n,t)\cdot\mathbf{f}_{a'}(n',t')} = 6\zeta\kBT\,\delta_{aa'}\,\delta(n-n')\,\delta(t-t').
\label{eq:fdt}
\end{equation}
The arms are coupled only through four boundary conditions (supplementary material, Sec.~A1),
\begin{equation}
\begin{aligned}
&\text{(i)}\;\; \mathbf{R}_1(0)=\mathbf{R}_2(0)=\mathbf{R}_3(0)=\mathbf{R}_4(0),\\
&\text{(ii)}\;\; {\textstyle\sum_a}\,\partial_n\mathbf{R}_a(0)=0,\\
&\text{(iii)}\;\; \mathbf{R}_1(N_L)=\mathbf{R}_2(N_L),\\
&\phantom{\text{(iii)}\;\;} \partial_n\mathbf{R}_1(N_L)+\partial_n\mathbf{R}_2(N_L)=0,\\
&\text{(iv)}\;\; \partial_n\mathbf{R}_3(N_3)=\partial_n\mathbf{R}_4(N_4)=0,
\end{aligned}
\label{eq:bc}
\end{equation}
which impose (i) continuity of the chain at the junction, (ii) force balance at the massless junction, (iii) closure of the loop with balanced tension at its far end, and (iv) force-free ends of the two tails.

We rescale each arm to the common domain $u=n/N_a\in[0,1]$ (Fig.~\ref{fig:schematic}), and measure lengths and times in units of the size and Rouse time of an unlooped chain of the same total length,
\begin{equation}
\hat{\mathbf{R}}_a = \frac{\mathbf{R}_a}{\sqrt{N_{tot}b^2}}, \qquad
\hat{t} = \frac{t}{\tau_R}, \qquad
\tau_R = \frac{\zeta N_{tot}^2b^2}{3\pi^2\kBT}.
\label{eq:scales}
\end{equation}
In these units, Eqs.~\eqref{eq:langevin} and \eqref{eq:fdt} keep their form in the arm coordinate $u$, with the stiffness and the noise strength of arm $a$ becoming $1/(\pi^2\hat n_a^2)$ and $2/(\pi^2\hat n_a)$, respectively (supplementary material, Sec.~A2). The topology therefore enters the equations of motion only through the arm lengths of Eq.~\eqref{eq:fractions}.

\subsection{Normal-mode decomposition and dynamics}
\label{sec:modes}

Because the arms have different lengths, the Rouse modes of a linear chain do not diagonalize Eq.~\eqref{eq:langevin}. Building on the exact eigenmodes of star and tadpole polymers~\cite{keesman2013}, which are mutually orthogonal only when all arms have equal length and the junction carries a correspondingly larger friction, we decompose the arm positions into a global translation $X$, a ``butterfly'' bending mode $Y$, an internal loop mode $Z_L$, and an internal tail mode $Z_T$ (supplementary material, Sec.~A3),
\begin{equation}
\begin{aligned}
\hat{\mathbf{R}}_1 &= X + \tfrac{1}{2}Z_L + \kappa Y, &\quad
\hat{\mathbf{R}}_2 &= X - \tfrac{1}{2}Z_L + \kappa Y, \\
\hat{\mathbf{R}}_3 &= X + \hat{n}_3 Z_T - Y, &\quad
\hat{\mathbf{R}}_4 &= X - \hat{n}_4 Z_T - Y.
\end{aligned}
\label{eq:modes}
\end{equation}
We write this map compactly as
\begin{equation}
\hat{\mathbf{R}} = \mathbf{V}\bm{\Psi}, \qquad
\bm{\Psi} = (X,Y,Z_L,Z_T)^{\mathrm{T}}, \qquad
\mathbf{P}\equiv\mathbf{V}^{-1},
\label{eq:VP}
\end{equation}
where $\mathbf{V}$ is a constant $4\times4$ matrix. The boundary conditions of Eq.~\eqref{eq:bc} transfer to two conditions on each mode, and the junction force balance decouples the boundary conditions of $X$ and $Y$ only for the amplitude ratio
\begin{equation}
\kappa = \frac{\hat{n}_L}{2}\left(\frac{1}{\hat{n}_3}+\frac{1}{\hat{n}_4}\right),
\label{eq:kappa}
\end{equation}
which is set by the topology alone. Left-multiplying the arm equations by $\mathbf{P}$ yields, without approximation, the equation of motion of the modes (supplementary material, Sec.~A4.1),
\begin{equation}
\begin{gathered}
\partial_{\hat{t}}\bm{\Psi} = \frac{1}{\pi^2}\,\bm{\Theta}\,\partial_u^2\bm{\Psi} + \mathbf{P}\hat{\mathbf{f}},
\qquad \bm{\Theta}\equiv\mathbf{P}\mathbf{D}\mathbf{V},\\
\mathbf{D}=\mathrm{diag}\!\left(\hat{n}_L^{-2},\hat{n}_L^{-2},\hat{n}_3^{-2},\hat{n}_4^{-2}\right).
\end{gathered}
\label{eq:mode_eom}
\end{equation}
The exchange symmetry of the two loop arms decouples $Z_L$ exactly, whereas $X$, $Y$, and $Z_T$ remain coupled through the non-symmetric matrix $\bm{\Theta}$ except when all arm lengths are equal~\cite{keesman2013}.

The boundary conditions select a different orthonormal basis for each mode (supplementary material, Sec.~A4.2),
\begin{equation}
\begin{gathered}
\phi^c_p = \sqrt{2}\cos(p\pi u), \qquad
\phi^s_p = \sqrt{2}\sin(p\pi u), \\
\phi^h_p = \sqrt{2}\sin\!\left[\left(p-\tfrac{1}{2}\right)\pi u\right],
\end{gathered}
\label{eq:bases}
\end{equation}
for $X$, $Z_L$, and the pair $\{Y,Z_T\}$, respectively, with eigenvalues $p^2$ for the first two families and $(p-\tfrac{1}{2})^2$ for the third. Expanding each mode in its basis up to $p_{\max}$ terms, $\Psi_i(u,\hat t)=\sum_{p=1}^{p_{\max}}\Psi_{i,p}(\hat t)\,\phi^{(i)}_p(u)$ (with, for $X$, an additional centre-of-mass amplitude $X_0$), and collecting the amplitudes $\Psi_{i,p}$ into a state vector $\mathcal{M}$ turns Eq.~\eqref{eq:mode_eom} into a multivariate Ornstein--Uhlenbeck process,
\begin{equation}
\frac{\diff\mathcal{M}}{\diff\hat{t}} = -\dmat{A}\,\mathcal{M} + \tilde{\mathbf{f}},
\qquad
\avg{\tilde{\mathbf{f}}(\hat{t})\,\tilde{\mathbf{f}}^{\mathrm{T}}(\hat{t}')} = \dmat{F}\,\delta(\hat{t}-\hat{t}').
\label{eq:OU}
\end{equation}
The position of any locus is linear in the state vector. Relative to the centre of mass $\hat{\mathbf{R}}_{cm}$ of the chain,
\begin{equation}
\begin{gathered}
\hat{\mathbf{R}}_a(u)-\hat{\mathbf{R}}_{cm}=\mathbf{B}_a(u)\,\mathcal{M},\\
\big[\mathbf{B}_a(u)\big]_{i,p}=V_{ai}\,\phi^{(i)}_p(u)-\sum_{b=1}^{4}\hat n_b V_{bi}\int_0^1\phi^{(i)}_p\,\diff u',
\end{gathered}
\label{eq:B}
\end{equation}
where $\mathbf{B}_a(u)$ is a row vector of length $4p_{\max}$: its entry for mode $i$ and index $p$ is the basis function of that mode at $u$, weighted by the entry $V_{ai}$ of $\mathbf{V}$, less its contour average over the whole chain, which removes the motion of the centre of mass (supplementary material, Sec.~A7.2). We write $\mathbf{B}(s)$ for the row of the locus at contour fraction $s$, which lies on arm $a$ at $s = s_a(1-u)$, $s_a + \hat n_L u$, $s_b - \hat n_L u$, or $s_b + \hat n_4 u$ for $a=3$, 1, 2, or 4, respectively (supplementary material, Sec.~A1.3).

The $p_{\max}\times p_{\max}$ blocks of the drift matrix $\dmat{A}$ and of the noise covariance matrix $\dmat{F}$ follow from the same matrix $\mathbf{P}$ (supplementary material, Secs.~A4.3--A5.2),
\begin{equation}
\begin{gathered}
\mathbf{A}_{ij} = \Theta_{ij}\,\bm{\Omega}_{ij}\,\bm{\lambda}^{(j)},
\qquad
\mathbf{F}_{ij} = \Gamma_{ij}\,\bm{\Omega}_{ij},\\
\bm{\Gamma}=\mathbf{P}\mathbf{g}\mathbf{P}^{\mathrm{T}},
\end{gathered}
\label{eq:AF}
\end{equation}
where $\bm{\lambda}^{(j)}$ holds the eigenvalues of mode $j$, $\mathbf{g}$ holds the arm noise strengths $2/(\pi^2\hat n_a)$, and $\bm{\Omega}_{ij}$ is the overlap between the bases of modes $i$ and $j$. This overlap is the identity when two modes share a basis, but couples the integer cosine basis of $X$ to the half-integer sine basis of $Y$ and $Z_T$.

This is the central technical point of the present theory. Allowing the loop and the tails to have arbitrary lengths places the modes on mutually non-orthogonal bases and makes $\bm{\Theta}$ non-symmetric. Consequently, unlike for a linear chain~\cite{doi1988,dutta2024,saini2025} or a symmetric star~\cite{keesman2013}, $\dmat{A}$ is not symmetric and has dense off-diagonal blocks. We accept this loss of orthogonality as the price of exactness: no approximation enters Eqs.~\eqref{eq:modes}--\eqref{eq:AF}, the model has no free parameter, and the topology enters only through the anchor fractions. 

\subsection{Equilibrium and time-dependent covariances}
\label{sec:corr}

The structure and dynamics of the chain follow from the equilibrium and time-dependent covariances of the mode amplitudes,
\begin{equation}
\mathbf{C}(0) = \avg{\mathcal{M}\,\mathcal{M}^{\mathrm{T}}}, \qquad
\mathbf{C}(\hat{t}) = \avg{\mathcal{M}(\hat{t})\,\mathcal{M}^{\mathrm{T}}(0)}.
\label{eq:Cdef}
\end{equation}
In the stationary state of Eq.~\eqref{eq:OU}, the equilibrium covariance is the unique symmetric, positive semi-definite solution of the continuous-time Lyapunov equation (supplementary material, Sec.~A6.1)
\begin{equation}
\dmat{A}\,\mathbf{C}(0) + \mathbf{C}(0)\,\dmat{A}^{\mathrm{T}} = \dmat{F},
\label{eq:lyapunov}
\end{equation}
which balances the covariance lost to drift against that injected by the noise.

In the supplementary material (Sec.~A6.2), we show that for the looped chain the solution of Eq.~\eqref{eq:lyapunov} is a diagonal matrix in closed form,
\begin{equation}
\mathbf{C}(0)=\frac{\dmat{K}^{-1}}{\pi^2},
\qquad
\avg{\Psi_{i,p}\Psi_{j,q}}=\frac{\delta_{ij}\,\delta_{pq}}{\pi^2\,k_i\,\lambda^{(i)}_p},
\label{eq:C0closed}
\end{equation}
where $\dmat{K}$ is diagonal with entries $k_i\lambda^{(i)}_p$. Here $\lambda^{(i)}_p$ are the eigenvalues of the basis functions of mode $i$, and $k_X=2/\hat n_L+1/\hat n_3+1/\hat n_4$, $k_Y=2\kappa^2/\hat n_L+1/\hat n_3+1/\hat n_4$, $k_{Z_L}=1/(2\hat n_L)$, and $k_{Z_T}=\hat n_3+\hat n_4$ are the elastic stiffnesses of the arms expressed in the modes. At equal times, the mode amplitudes are therefore independent. The coupling of the modes through $\dmat{F}$ affects how they relax, but not their equilibrium statistics. The equilibrium covariance thus follows analytically for any topology, with the anchor fractions entering only through the four stiffnesses $k_i$.

Since the noise at later times is uncorrelated with the present state, the time-dependent covariance relaxes under the same operator that governs the mean (supplementary material, Sec.~A6.3),
\begin{equation}
\mathbf{C}(\hat{t}) = e^{-\dmat{A}\hat{t}}\,\mathbf{C}(0), \qquad \hat{t}\geq0 .
\label{eq:Ct}
\end{equation}
Because $\dmat{A}$ is not diagonal, $\mathbf{C}(\hat t)$ is non-diagonal, the modes do not relax independently, and the loop couples the relaxation of the translation, bending, and tail modes. Together, Eqs.~\eqref{eq:C0closed} and \eqref{eq:Ct} determine every structural and dynamical observable of the looped chain, including its radius of gyration and the mean squared displacement of any locus, as functions of the anchor fractions alone (supplementary material, Secs.~A7 and A8). 
\section{Results}
\label{sec:results}

\begin{figure*}[t]
  \centering
  \includegraphics{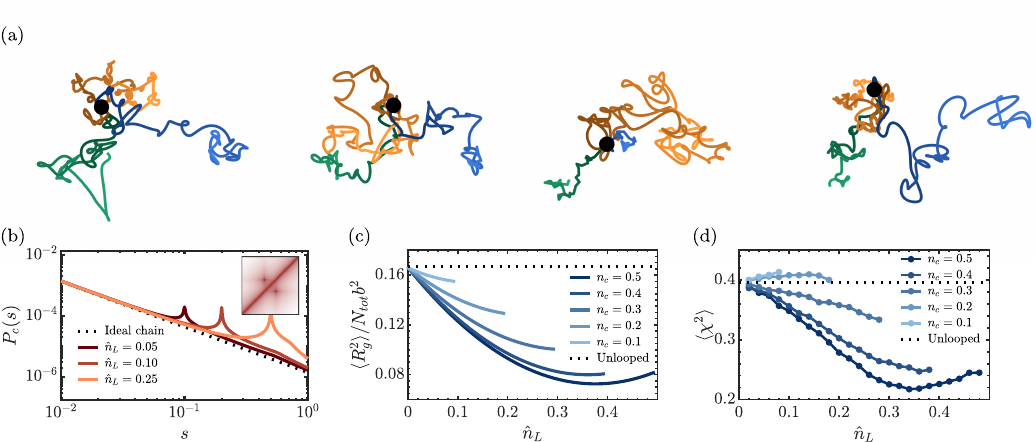}
  \caption{Equilibrium structure of a Rouse chain with a single loop: (a)~Representative configurations from the equilibrium distribution for chains with a loop centred on the chain ($n_c = 0.5$) with loop size $\hat n_L = 0.15$, $0.30$, and $0.45$, and of a chain with an off-centre loop ($n_c = 0.3$, $\hat n_L = 0.15$), from left to right. The loop is shown in orange and the two tails in green and blue.  (b)~Contact probability $P_c(s)$ as a function of the contour separation $s$ for an ideal chain (dotted) and for centred loops with $\hat n_L = 0.05$, $0.10$, and $0.25$. Inset: contact map $\log_{10} P_c(|\Delta\mathbf{R}| < r_c)$ for $\hat n_L = 0.10$. The capture radius is $r_c = 0.01\sqrt{N_{tot}}\,b$.
  (c)~Mean squared radius of gyration and (d)~mean relative shape anisotropy as functions of $\hat n_L$ for loop centres from $n_c = 0.5$ (dark) to $0.1$ (light). Dotted lines: unlooped Rouse chain. Each point in (d) averages $10^4$ configurations.}
  \label{fig:structure}
\end{figure*}

\subsection{Equilibrium structure}
\label{sec:structure}

Because the equations of motion are linear and driven by Gaussian thermal noise, the stationary distribution of the mode amplitudes is a Gaussian with zero mean and covariance $\mathbf{C}(0)$, Eq.~\eqref{eq:C0closed}. The equilibrium covariance therefore contains the complete equilibrium statistics of the looped chain. For a given topology, i.e., a given pair of anchor fractions $(s_a, s_b)$, we characterize the loop by its size $\hat n_L$ and by the position of its centre along the chain, $n_c = (s_a + s_b)/2$.
Sampling the mode amplitudes from this Gaussian~\cite{dutta2024,saini2025} and transforming back to the arm positions yields statistically exact equilibrium configurations for any topology (supplementary material, Sec.~B2). Representative configurations are shown in Fig.~\ref{fig:structure}(a). 

The equilibrium covariance also determines the spatial proximity of any two loci along the chain. For two loci at contour fractions $s_1$ and $s_2$, the separation vector $\Delta\mathbf{R}$ is a linear combination of the mode amplitudes and is therefore an isotropic Gaussian vector. Its mean square follows directly from $\mathbf{C}(0)$,
\begin{equation}
\avg{|\Delta\mathbf{R}|^2} = \left[\mathbf{B}(s_1) - \mathbf{B}(s_2)\right]\mathbf{C}(0)\left[\mathbf{B}(s_1) - \mathbf{B}(s_2)\right]^{\mathrm{T}},
\label{eq:dR2}
\end{equation}
an explicit sum over the modes by Eq.~\eqref{eq:C0closed}, where $\mathbf{B}(s)$ is the projection of Eq.~\eqref{eq:B}. The magnitude $|\Delta\mathbf{R}|$ therefore follows a Maxwell distribution, and the probability that the two loci are closer than a capture radius $r_c$ follows exactly,
\begin{equation}
P_c(|\Delta\mathbf{R}| < r_c) = \operatorname{erf}(x) - \frac{2}{\sqrt{\pi}}\,x\,e^{-x^2},
\quad
x = r_c\sqrt{\frac{3}{2\avg{|\Delta\mathbf{R}|^2}}},
\label{eq:contact}
\end{equation}
which yields a contact map analogous to a Hi-C map [Fig.~\ref{fig:structure}(b), inset]. The loop appears as a dot at the anchors, where the two anchors are in contact with certainty, together with enhanced contacts within the loop and between loci flanking the anchors. Averaging Eq.~\eqref{eq:contact} over all pairs of loci at the same contour separation $s$, on a uniform grid of $1000$ loci, gives the contact probability $P_c(s)$ [Fig.~\ref{fig:structure}(b)]. For an ideal chain, $P_c(s) \propto s^{-3/2}$ for $r_c \ll \sqrt{N_{tot}b^2}$. With a loop, $P_c(s)$ follows the ideal scaling at small $s$ but rises above it as $s$ approaches the contour length of the loop, $2\hat n_L$, where it peaks because pairs of loci flanking the anchors are brought into contact. For $s > 2\hat n_L$, $P_c(s)$ does not recover the ideal curve: the loop acts as a shortcut between loci on either side of it, and contacts remain enhanced at all larger separations and the enhancement grows with the loop size.%

Beyond pairwise distances, the overall size of the chain is quantified by the mean squared radius of gyration, the contour-weighted mean square distance of the loci from the centre of mass. Because $\mathbf{C}(0)$ is diagonal, it reduces to a single sum over the modes, in units of $N_{tot}b^2$,
\begin{equation}
\begin{gathered}
\avg{\hat R_g^2}=\sum_{i}\sum_{p}C_{i,p}\,G_{i,p},\\
G_{i,p}=\sum_{a=1}^{4}\hat n_a\int_0^1\big[\mathbf{B}_a(u)\big]_{i,p}^2\,\diff u,
\end{gathered}
\label{eq:Rg}
\end{equation}
where $C_{i,p}=1/(\pi^2k_i\lambda^{(i)}_p)$ are the diagonal entries of $\mathbf{C}(0)$, Eq.~\eqref{eq:C0closed}, and the geometric factors $G_{i,p}$ are elementary integrals of the basis functions; the sums over $p$ are carried out analytically (supplementary material, Sec.~A7.2). As expected, a loop of any size or position compacts the chain, and $\avg{R_g^2}$ lies below the value $N_{tot}b^2/6$ of the unlooped chain for all topologies [Fig.~\ref{fig:structure}(c)]. In the limits $\hat n_L \to 0$ and $\hat n_L \to 1/2$, it approaches the values of an ideal linear chain, $N_{tot}b^2/6$, and of an ideal ring, $N_{tot}b^2/12$, respectively~\cite{doi1988}. Interestingly, $\avg{R_g^2}$ is not monotonic in the loop size: for a centred loop it reaches a minimum of $\approx 0.073\,N_{tot}b^2$ at $\hat n_L \approx 0.37$, below the ring value, so that a large loop flanked by two short tails is more compact than a ring of the same total length. This effect weakens as the loop moves away from the centre of the chain, and for $n_c \lesssim 0.3$, $\avg{R_g^2}$ decreases monotonically up to the largest loop that fits.

Finally, we quantify the shape of the chain by the relative shape anisotropy~\cite{theodorou1985,aronovitz1986,wei1990,saini2025}
\begin{equation}
\chi^2 = \frac{3}{2}\,\frac{\Lambda_1^2 + \Lambda_2^2 + \Lambda_3^2}{(\Lambda_1 + \Lambda_2 + \Lambda_3)^2} - \frac{1}{2},
\label{eq:aniso}
\end{equation}
where $\Lambda_1$, $\Lambda_2$, and $\Lambda_3$ are the eigenvalues of the gyration tensor of a configuration; $\chi^2 = 0$ for a sphere and $1$ for a rod. Since $\chi^2$ is a nonlinear function of each configuration, we average it over configurations sampled from $\mathbf{C}(0)$. Like $\avg{R_g^2}$, $\avg{\chi^2}$ approaches the values of the ideal linear chain ($\approx 0.40$~\cite{wei1990}) and of the ideal ring ($\approx 0.25$) for $\hat n_L \to 0$ and $\hat n_L \to 1/2$, and is non-monotonic for a centred loop, with a minimum of $\approx 0.22$ at $\hat n_L \approx 0.34$ [Fig.~\ref{fig:structure}(d)]. The chain becomes increasingly anisotropic as the loop moves off-centre. Remarkably, when the loop lies close to one end of the chain ($n_c = 0.1$), the chain is more anisotropic than an unlooped chain, with $\avg{\chi^2} \approx 0.41$ against $0.396$: a compact loop at one end, attached to a long tail, elongates the chain.

\subsection{Dynamics of a tagged locus}
\label{sec:dynamics}

\begin{figure*}[t]
  \centering
  \includegraphics{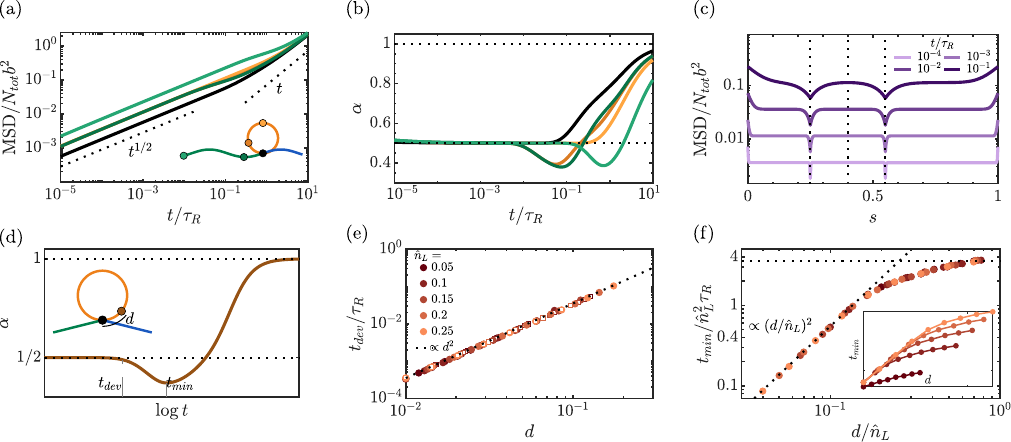}
  \caption{Dynamics of tagged loci on a Rouse chain with a single loop: (a)~Mean squared displacement (MSD) and (b)~local exponent $\alpha$ of five loci on a chain with anchors at $s_a = 0.25$ and $s_b = 0.55$: the loop centre, the junction, a locus on the loop, a locus on the first tail, and the free end of the first tail. The sketch in (a) shows their relative positions, in the colours of the corresponding curves. Dotted lines: $t^{1/2}$ and $t$ in (a), $\alpha = 1/2$ and $1$ in (b). (c)~MSD as a function of the contour position $s$ along the chain at $t/\tau_R = 10^{-4}$ to $10^{-1}$ (light to dark). Dotted vertical lines represent the two anchors and the loop centre. (d)~Schematic of the dip in $\alpha$ for a locus at contour distance $d$ from the junction, defining the onset time $t_{dev}$, at which $\alpha$ first falls below $1/2$, and the time $t_{min}$ of the minimum of $\alpha$.
  (e)~$t_{dev}$ as a function of $d$ for loci on the loop (filled circles) and on a tail (open squares) of chains with a centred loop, $\hat n_L = 0.05$ (dark) to $0.25$ (light). Dotted line represents a line with $t_{dev} \propto d^2$.
  (f)~$t_{min}$ of loci on the loop, scaled as $t_{min}/\hat n_L^2$ against $d/\hat n_L$, for the same loops. Dotted lines represent the $\propto (d/\hat n_L)^2$ and a constant representing the saturation. The same values without the proposed scaling in either axis are shown in the inset.}
  \label{fig:dynamics}
\end{figure*}

The time-dependent covariance $\mathbf{C}(\hat t)$, Eq.~\eqref{eq:Ct}, gives access to the dynamics of the looped chain in the same analytical manner. We focus on the mean squared displacement (MSD) of a tagged locus, the quantity most commonly measured in live-cell imaging of chromatin~\cite{weber2010,hajjoul2013,gabriele2022,mach2022}. The centre of mass of the chain performs free Brownian motion, driven by noise that is uncorrelated with the internal modes, so the displacement of a locus splits into a centre-of-mass part and an internal part without cross terms (supplementary material, Sec.~A8.1). The internal position of the locus at contour fraction $s$ is the projection $\mathbf{B}(s)\mathcal{M}$ of Eq.~\eqref{eq:B}. Combining it with Eq.~\eqref{eq:Ct} yields the MSD (supplementary material, Sec.~A8.3),
\begin{equation}
\avg{\Delta\hat{\mathbf{R}}^2(s,\hat t)} = \frac{2}{\pi^2}\,\hat t
  + 2\,\mathbf{B}(s)\left[\mathbf{I}-e^{-\dmat{A}\hat t}\right]\mathbf{C}(0)\,\mathbf{B}^{\mathrm{T}}(s),
\label{eq:msd}
\end{equation}
as a function of the time lag, the position of the locus, and the two anchor fractions, with no adjustable parameter. We characterize its time dependence by the local exponent
\begin{equation}
\alpha(\hat t) = \frac{\diff\ln\avg{\Delta\hat{\mathbf{R}}^2}}{\diff\ln\hat t},
\label{eq:alpha}
\end{equation}
which we evaluate analytically from the time derivative of Eq.~\eqref{eq:msd}. For an unlooped Rouse chain, $\alpha$ increases from the subdiffusive value $1/2$ at short times to $1$ at long times, when the whole chain diffuses~\cite{doi1988}.

Figure~\ref{fig:dynamics}(a) shows the MSD of five loci on a chain with an off-centre loop. All follow $t^{1/2}$ at short times, with different prefactors, and free centre-of-mass diffusion, $\propto t$, at long times. Along the chain, the junction is the slowest locus: at short times its MSD is half that of a locus in the interior of an arm, because it is attached to four strands instead of two. The free ends of the tails are the fastest, with twice the interior value, because they are attached to a single strand. Within the loop, the loop centre is the most mobile locus at all times. The MSD along the whole chain [Fig.~\ref{fig:dynamics}(c)] shows how the influence of the anchors spreads with time. At short time lags, the slowing is localized at the anchors but this region widens as the time lag progresses, as the anchors constrain progressively longer stretches of the chain.

The local exponent reveals a qualitative departure from Rouse behaviour [Fig.~\ref{fig:dynamics}(b)]. For loci near the anchor, on the loop as well as on a tail, $\alpha$ first dips below $1/2$ before it rises to $1$; the free end of the tail shows the same dip at later times. The only loci without a dip are the junction and the loop centre. These features agree with bead--spring simulations of rosette-like loop configurations, where loci show a stretching exponent well below $1/2$ at intermediate times~\cite{yuan2024}. The dip has a simple origin. Any locus initially moves together with the stretch of chain around it, which grows diffusively along the two strands that meet at the locus. When this stretch reaches the junction, it spreads into the other three strands, which slows the locus further. We characterize the dip by two times [Fig.~\ref{fig:dynamics}(d)]: the time $t_{dev}$ at which $\alpha$ first falls below $1/2$, which marks the deviation from Rouse behaviour, and the time $t_{min}$ at which $\alpha$ is minimal. To determine how these times depend on the architecture, we computed them for loci at contour distance $d$ (in units of $N_{tot}$) from the junction, on the loop and on a tail, for centred loops of sizes $\hat n_L = 0.05$ to $0.25$ (supplementary material, Sec.~B3). The onset time is independent of the loop size and of the side of the junction on which the locus lies: for all loops, $t_{dev} \propto d^2$ [Fig.~\ref{fig:dynamics}(e)], with $t_{dev} \approx 3.4\,d^2\tau_R$ (supplementary material, Sec.~B3). Since $d^2\tau_R$ is the Rouse time of the segment of $dN_{tot}$ monomers between the locus and the anchor, $t_{dev}$ is the time required for stress to be communicated from the anchor to the locus along the chain.

The time of the minimum depends on both $d$ and $\hat n_L$ [Fig.~\ref{fig:dynamics}(f), inset]. Nevertheless, the data for all loop sizes collapse onto a single curve when $t_{min}/\hat n_L^2$ is plotted against $d/\hat n_L$ [Fig.~\ref{fig:dynamics}(f)]. For loci close to the anchor, $d \ll \hat n_L$, $t_{min} \propto d^2$ ($\approx 54\,d^2\tau_R$; supplementary material, Sec.~B3) and is independent of the loop size. As the locus approaches the loop centre, the dependence on $d$ weakens and $t_{min}$ approaches $\approx 3.6\,\hat n_L^2\tau_R$ [Fig.~\ref{fig:dynamics}(f)], of the order of the Rouse time of a loop arm. Here the dip ends earlier, when stress has propagated around the entire loop and no fresh chain within the loop remains to be recruited. The minimum is therefore set by the shorter of two times: the time for the locus to join the motion of the anchor, controlled by $d$, and the time for stress to traverse the loop, controlled by $\hat n_L$.

Altogether, the motion of an individual locus carries a qualitative signature that it belongs to an architectural loop, a transient dip of $\alpha$ below the Rouse value, and quantitative signatures of the specific architecture.

\subsection{Inferring the loop anchors from short-time motion}
\label{sec:inference}

The dynamics of a tagged locus carries the signature of the loop architecture. Reading this signature from the dip in $\alpha$, however, requires tracking the locus over several decades in time, which may not be experimentally realistic. Moreover, converting $t_{dev}$ and $t_{min}$ into contour distances requires the Rouse time $\tau_R$ of the chain. Polymer models have been used to reconstruct the connectivity of chromatin from live images of several tagged loci~\cite{dutta2023}. We therefore ask whether a simpler dynamical observable, accessible at short times alone, can reveal the positions of the anchors.

\begin{figure}[t]
  \centering
  \includegraphics{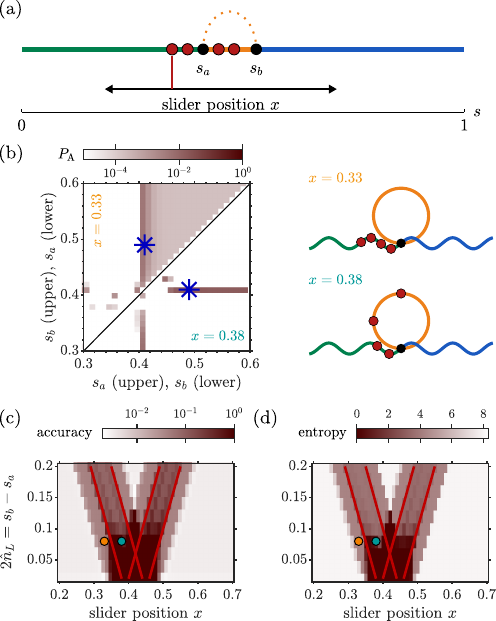}
  \caption{Inference of the loop anchors from the short-time motion of four tagged loci:
  (a)~Four loci (red circles) at contour positions $x$, $x+0.01$, $x+0.04$, and $x+0.06$ slide along a chain with a loop between the anchors $s_a$ and $s_b$ (black circles); the slider position $x$ is the position of the first locus. Not to scale.
  (b)~Probability $P_\mathrm{A}(s_a, s_b)$ of the anchor positions for a loop with $s_a = 0.41$ and $s_b = 0.49$ (asterisks) at two slider positions: $x = 0.33$ (upper triangle) and $x = 0.38$ (lower triangle, mirrored across the diagonal). The sketches on the right show the positions of the four loci relative to the loop for the two slider positions (not to scale).
  (c)~Accuracy, the probability assigned to anchors within $\pm 0.01$ of the true positions, and (d)~entropy of $P_\mathrm{A}$ as functions of the slider position $x$ and the loop length $2\hat n_L = s_b - s_a$, for loops centred at $n_c = 0.45$. Both colour scales run from the value of a random guess (white) to a correct and certain prediction (dark red). Red lines: slider positions at which the first or the last locus sits on an anchor. The orange and cyan circles represent the two configurations shown in the upper and lower triangles, respectively, in (b).}
  \label{fig:inference}
\end{figure}

At short times, $t \ll \tau_R$, every locus moves subdiffusively with $\alpha \approx 1/2$ [Fig.~\ref{fig:dynamics}(b)], and its MSD differs from that of a locus in the interior of an arm only by a prefactor set by its connectivity: half at the junction, twice at a free end, and intermediate values near them [Fig.~\ref{fig:dynamics}(c)]. The ratio of the MSDs of two loci is therefore nearly independent of time in this regime but depends on where the loci lie relative to the anchors. To test whether such ratios suffice to locate a loop, we consider four tagged loci at contour positions $s_1 = x$, $s_2 = x + 0.01$, $s_3 = x + 0.04$, and $s_4 = x + 0.06$, and slide this group along the chain by varying $x$ [Fig.~\ref{fig:inference}(a)]. The unequal spacings make all six pairwise separations different. The MSDs of four loci yield three independent ratios, $r_k(\hat t) = \avg{\Delta\hat{\mathbf{R}}^2(s_k,\hat t)}/\avg{\Delta\hat{\mathbf{R}}^2(s_4,\hat t)}$ for $k = 1, 2, 3$. We compute them from Eq.~\eqref{eq:msd} at $n_t = 6$ logarithmically spaced times $\hat t_i$ between $10^{-4}$ and $10^{-2}$ and average their logarithms over this window,
\begin{equation}
g_k = \frac{1}{n_t}\sum_{i=1}^{n_t} \ln r_k(\hat t_i).
\label{eq:gk}
\end{equation}
We evaluate $g_k$ for every topology with both anchors on a grid of spacing $0.01$ between $0.05$ and $0.95$, which gives $N = 4095$ candidate pairs $(s_a, s_b)$ with $s_a < s_b$ (supplementary material, Sec.~B4).

To quantify how much information the ratios carry about the anchors, we take the measured values $g_k^*$ to be those of the true topology, $(s_a^*, s_b^*)$, without added noise (supplementary material, Sec.~B4). With a uniform prior over the candidates, the probability that the anchors lie at $(s_a, s_b)$ is
\begin{equation}
P_\mathrm{A}(s_a, s_b) = \frac{e^{-\mathcal{D}^2(s_a, s_b)/2}}{\sum_{s_a' < s_b'} e^{-\mathcal{D}^2(s_a', s_b')/2}},
\label{eq:PA}
\end{equation}
where
\begin{equation}
\mathcal{D}^2(s_a, s_b) = \sum_{k=1}^{3}\frac{\left[g_k^* - g_k(s_a, s_b)\right]^2}{\sigma_g^2}
\label{eq:chi2}
\end{equation}
measures the deviation of the ratios of a candidate topology from the measured ones, in units of a tolerance $\sigma_g=0.002$ (supplementary material, Sec.~B4). We characterize $P_\mathrm{A}$ by two measures. The accuracy, $\mathcal{A}$, is the total probability that $P_\mathrm{A}$ assigns to anchors within one grid spacing of the true ones, $|s_a - s_a^*| \le 0.01$ and $|s_b - s_b^*| \le 0.01$. The entropy, $S = -\sum P_\mathrm{A}\ln P_\mathrm{A}$, measures the certainty of the prediction. The two measures are complementary, since a prediction can be certain and yet wrong. A random guess gives $\mathcal{A} = 9/N \approx 0.002$ and $S = \ln N \approx 8.3$, whereas a correct and certain prediction gives $\mathcal{A} = 1$ and $S = 0$.

Figure~\ref{fig:inference}(b) shows $P_\mathrm{A}$ for a loop with $s_a^* = 0.41$ and $s_b^* = 0.49$ at two slider positions. At $x = 0.38$, two loci lie on the tail and two within the loop, so that the group straddles the anchor $s_a$. The probability is then concentrated in a compact region around the true anchors, with $\mathcal{A} = 0.97$ and $S \approx 0.9$. At $x = 0.33$, all four loci lie on the tail, between $0.02$ and $0.08$ from the anchor. The probability still favours anchors near the true $s_a$ but spreads over a broad range of $s_b$, with $\mathcal{A} = 0.04$ and $S \approx 4.3$.

Figures~\ref{fig:inference}(c) and \ref{fig:inference}(d) show the accuracy and the entropy for loops centred at $n_c = 0.45$ with $2\hat n_L = 0.02$ to $0.20$ and for slider positions from $x = 0.2$ to $0.7$. Good predictions are confined to two V-shaped bands, bounded by the slider positions at which the first or the last locus sits on an anchor. Within a band, the group of loci overlaps an anchor. Outside the bands, the accuracy and the entropy do not differ from those of a random guess. This holds for loci on the tails as well as for loci deep inside a large loop. For small loops, $2\hat n_L \le 0.06$, the two bands merge and the group can span both anchors. For larger loops, the bands separate and the group overlaps only one anchor at a time. Altogether, the short-time MSD ratios of a few loci placed around one of the anchors identify the anchors accurately and with high certainty, without long trajectories and without knowledge of the Rouse time beyond the requirement that the observation window lie in the short-time regime.

\section{Discussion and conclusions}
\label{sec:discussion}

In this manuscript, we extended the Rouse theory of flexible polymers to a chain with a single loop of arbitrary size and position. The key step is a modal decomposition of the looped chain, viewed as a four-arm star with arms of unequal lengths, that respects the boundary conditions at the junction exactly. The resulting modes are not orthogonal, but their equilibrium covariance follows in closed form and their time-dependent covariance exactly, and together they determine the structure and dynamics of the chain as functions of the two anchor positions alone, with no adjustable parameter. Because the theory is analytical, it gives these properties for any topology and over many decades in time, from the motion of short chain segments to the free diffusion of the whole chain. Such a range is difficult to resolve with live-cell imaging, which typically follows a locus over one or two decades in time, and costly to cover with simulations.

The framework recovers the expected structural effects of a loop: it compacts the chain and enhances long-range contacts. Its main results concern dynamics. Over long times, the motion of a single locus carries the signature of the loop architecture: the local exponent of its MSD dips below the Rouse value, and the onset of the dip measures the contour distance of the locus from the nearest anchor, and its minimum also reflects the size of the loop. Over short times, all loci move with the same exponent, but the ratios of their MSDs depend on where they lie relative to the anchors. The MSD ratios of a few loci placed around an anchor are then predictive of the loop architecture: a Bayesian analysis of these ratios locates the anchors accurately and with high certainty, without long trajectories.

Several extensions follow naturally. First, chromatin contains many loops, which may be nested or arranged in series~\cite{fudenberg2016,polovnikov2023}, and the modal decomposition introduced here can be generalized to multiple loops and more complex topologies, complementing numerical studies of chromatin with many loops and dynamic loop extrusion~\cite{yuan2024}; how the dynamical signatures of neighbouring loops interact remains to be understood. Second, loops in interphase chromatin are formed by active extrusion~\cite{sanborn2015,fudenberg2016,nuebler2018}, so that the topological constraint of a loop is accompanied by active forces on the chain~\cite{ghosh2022,dutta2024}, and the anchors themselves move as the loop grows. Combining the present framework with such forces would show how the constraint and the activity together shape the dynamics of tagged loci. Third, the predictions should be tested against live-cell measurements of loci with known loop anchors~\cite{gabriele2022,mach2022,bruckner2023}. Such a comparison would also show how far the Rouse description, which neglects hydrodynamic and excluded-volume interactions, can be pushed in interpreting the motion of chromatin loci. Altogether, the present theory adds loop architecture to the toolbox of the theory of flexible polymers and brings the theory a step closer to describing the dynamics of biopolymers.

\section*{Supplementary Material}
The supplementary material contains, in Part A, the full derivation of the looped-chain model: the physical model and its non-dimensionalization, the normal-mode coordinates, the drift and noise covariance matrices, the equilibrium and time-dependent covariances, the radius of gyration, and the mean squared displacement. Part B describes the numerical methods: the implementation, the sampling of equilibrium configurations, the analysis of the dip in the local exponent, and the inference of the loop anchors.

\begin{acknowledgments}
We thank Ashesh Ghosh for helpful discussions. S.D. acknowledges the IIT Bombay seed grant (Project No. RD/0523-IRCCSH0-022) and the PMECRGA grant from the Anusandhan National Research Foundation (ANRF), India, vide file no.\ ANRF/ECRG/2024/001067/ENS.
\end{acknowledgments}

\section*{Author Declarations}
\subsection*{Conflict of Interest}
The authors have no conflicts to disclose.

\subsection*{Author Contributions}

\textbf{Rishan Gobse}: Formal analysis (equal); Methodology (equal).
\textbf{Rishit Kesharwani}: Formal analysis (equal); Methodology (equal).
\textbf{Sayantan Dutta}: Conceptualization (lead); Formal analysis (equal); Methodology (equal); Funding acquisition (lead); Writing -- original draft (lead).

\subsection*{Use of Artificial Intelligence Tools}
The authors used the large language model Claude Opus 5.5 (Anthropic PBC) to verify and organize the derivations developed by the authors, to rephrase the text of the manuscript, and to write MATLAB functions that implement the expressions strictly as defined in this manuscript, all under the authors' direction. The tool was used to improve readability. The authors verified all outputs generated with this tool and take full responsibility for the content of this article.

\section*{Data Availability}
All calculations follow the equations given in the article and its supplementary material. The raw data of the calculations are available from the corresponding author upon reasonable request.

\clearpage
\onecolumngrid
\setcounter{equation}{0}
\renewcommand{\theequation}{S\arabic{equation}}
\begin{center}
{\large\bfseries Supplementary Material: Analytical Rouse theory reveals how a loop shapes the structure and dynamics of a flexible polymer}\\[1em]
Rishan Gobse, Rishit Kesharwani, and Sayantan Dutta
\end{center}

\sipart{Theory}
\label{S:part:theory}

\sinote{The physical model}

The molecule is a four-arm star with two arms closed into a loop and two free tails, joined at a common junction. With $N_L$ the length of each loop arm and $N_3,N_4$ the tail lengths, the total contour length is
\begin{equation}
    N_{tot}=2N_L+N_3+N_4 .
\end{equation}
Each arm is a continuous Gaussian thread: $\mathbf R_a(n,t)$, $a=1,2,3,4$, is the position of contour point $n\in[0,N_a]$ on arm $a$ at time $t$, with $n=0$ the junction.

\sisub{Governing equation}
The Kuhn length $b$ is the same for every segment, so that a bond of unit contour length has mean-square extension $b^2$ and the entropic spring constant per unit contour length is $k=3k_BT/b^2$. With viscous drag coefficient $\zeta$ per unit contour length, each arm obeys
\begin{equation}
    \zeta\,\partial_t \mathbf R_a(n,t) = k\,\partial_n^2 \mathbf R_a(n,t) + \mathbf f_a(n,t),
    \label{S:eq:rouse_dim}
\end{equation}
with thermal noise satisfying the fluctuation--dissipation theorem, written for the dot product of the three-dimensional vector noise,
\begin{equation}
    \langle \mathbf f_a(n,t)\cdot \mathbf f_{a'}(n',t')\rangle = 6\zeta k_BT\,\delta_{aa'}\,\delta(n-n')\,\delta(t-t').
    \label{S:eq:fdt_dim}
\end{equation}

\sisub{Boundary conditions}
\begin{enumerate}
    \item[(B1)] Junction continuity: $\mathbf R_1(0,t)=\mathbf R_2(0,t)=\mathbf R_3(0,t)=\mathbf R_4(0,t)$.
    \item[(B2)] Junction force balance (massless node): $\displaystyle\sum_{a=1}^4 \partial_n\mathbf R_a(0,t)=0$.
    \item[(B3)] Loop closure and tension balance: $\mathbf R_1(N_L,t)=\mathbf R_2(N_L,t)$ and $\partial_n\mathbf R_1(N_L,t)+\partial_n\mathbf R_2(N_L,t)=0$.
    \item[(B4)] Free tail ends: $\partial_n\mathbf R_3(N_3,t)=\partial_n\mathbf R_4(N_4,t)=0$.
\end{enumerate}

\sisub{Fractional contour coordinates}
Each arm is rescaled to the common domain $u=n/N_a\in[0,1]$, with fractional lengths
\begin{equation}
    \hn L=\frac{N_L}{N_{tot}},\qquad \hn3=\frac{N_3}{N_{tot}},\qquad \hn4=\frac{N_4}{N_{tot}},\qquad 2\hn L+\hn3+\hn4=1,
    \label{S:eq:fracs}
\end{equation}
parametrized by two anchor fractions $s_a<s_b\in[0,1]$ through $\hn3=s_a$, $\hn L=(s_b-s_a)/2$, $\hn4=1-s_b$. A locus at contour fraction $s$ along the whole chain lies on arm $a$ at the rescaled coordinate $u$ given by
\begin{equation}
    s=\begin{cases}
        s_a(1-u), & a=3,\\
        s_a+\hn L\,u, & a=1,\\
        s_b-\hn L\,u, & a=2,\\
        s_b+\hn4\,u, & a=4,
    \end{cases}
    \label{S:eq:s_to_u}
\end{equation}
so that both anchors, $s=s_a$ and $s=s_b$, map to the junction, $u=0$. Equation~\eqref{S:eq:rouse_dim} becomes
\begin{equation}
    \zeta\,\partial_t \mathbf R_a(u,t) = \frac{k}{N_a^2}\,\partial_u^2 \mathbf R_a(u,t) + \mathbf f_a(u,t),
    \qquad
    \langle \mathbf f_a\!\cdot\! \mathbf f_{a'}\rangle = \frac{6\zeta k_BT}{N_a}\,\delta_{aa'}\delta(u-u')\delta(t-t').
    \label{S:eq:rouse_u}
\end{equation}

\sinote{Non-dimensionalization}

Equation~\eqref{S:eq:rouse_u} contains one length scale and one time scale. Both are removed before any internal coordinate is introduced, using the equilibrium coil size and the Rouse time of the unlooped reference chain of the same total contour length.
\begin{equation}
    \hR_a = \frac{\mathbf R_a}{\sqrt{N_{tot}b^2}}, \qquad
    \hat t=\frac{t}{\tau_R}, \qquad
    \tau_R=\frac{\zeta N_{tot}^2b^2}{3\pi^2k_BT}=\frac{\zeta N_{tot}^2}{\pi^2 k}.
    \label{S:eq:scales}
\end{equation}

This scaling is independent of the anchor positions. Substituting \eqref{S:eq:scales} into \eqref{S:eq:rouse_u} and using $k\tau_R/\zeta=N_{tot}^2/\pi^2$,
\begin{equation}
    \partial_{\hat t} \hR_a(u,\hat t) = \frac{1}{\pi^2\hn a^2}\,\partial_u^2 \hR_a(u,\hat t) + \hf_a(u,\hat t)\
    \label{S:eq:rouse_dimless}
\end{equation}
with the rescaled noise $\hf_a=\tau_R\,\mathbf f_a/(\zeta\sqrt{N_{tot}b^2})$ obeying
\begin{equation}
    \langle \hf_a(u,\hat t)\cdot \hf_{a'}(u',\hat t')\rangle = \frac{2}{\pi^2\hn a}\,\delta_{aa'}\,\delta(u-u')\,\delta(\hat t-\hat t')\
    \label{S:eq:fdt_dimless}
\end{equation}
The prefactor $6$ of \eqref{S:eq:fdt_dim} is reduced to $2$ by $\tau_R/(\zeta N_{tot}b^2)=N_{tot}/(3\pi^2k_BT)$. Equations~\eqref{S:eq:rouse_dimless}--\eqref{S:eq:fdt_dimless} contain the entire physical content of the model; $N_{tot}$, $\zeta$, $k_BT$ and $b$ do not appear again.

Collecting the four arms into a single column,
\begin{equation}
    \partial_{\hat t}\hR=\frac{1}{\pi^2}\,\mathbf D\,\partial_u^2\hR+\hf,
    \qquad
    \hR\equiv\begin{pmatrix}\hR_1\\ \hR_2\\ \hR_3\\ \hR_4\end{pmatrix},
    \qquad
    \mathbf D\equiv\mathrm{diag}\!\left(\frac{1}{\hn L^{2}},\;\frac{1}{\hn L^{2}},\;\frac{1}{\hn3^{2}},\;\frac{1}{\hn4^{2}}\right),
    \label{S:eq:arm_matrix}
\end{equation}
using $\hn1=\hn2=\hn L$.

\sinote{Normal-mode coordinates}
\label{S:sec:modes}

\sisub{Definitions and the transformation matrix}
The arm positions are decomposed into a global translation $X$, a ``butterfly'' bending mode $Y$, an internal loop mode $Z_L$, and an internal tail mode $Z_T$:
\begin{equation}
\begin{aligned}
    \hR_1 &= X + \tfrac12 Z_L + \kappa Y, &\qquad
    \hR_2 &= X - \tfrac12 Z_L + \kappa Y, \\
    \hR_3 &= X + \hn3 Z_T - Y, &\qquad
    \hR_4 &= X - \hn4 Z_T - Y.
\end{aligned}
\label{S:eq:Rdef}
\end{equation}
In Sec.~\ref{S:sec:bc_alpha}, we will show that  $\kappa$ is uniquely determined by the boundary conditions of the problem.

Collecting the modes into a column $\Ps\equiv(X,\,Y,\,Z_L,\,Z_T)^T$, Eq.~\eqref{S:eq:Rdef} is the linear map
\begin{equation}
    \hR=\mathbf V\,\Ps,
    \qquad
    \mathbf V=
    \begin{pmatrix}
        1 & \kappa & \tfrac12 & 0\\[3pt]
        1 & \kappa & -\tfrac12 & 0\\[3pt]
        1 & -1 & 0 & \hn3\\[3pt]
        1 & -1 & 0 & -\hn4
    \end{pmatrix},
    \label{S:eq:Vmatrix}
\end{equation}
whose inverse
\begin{equation}
    \Ps=\mathbf P\,\hR,
    \qquad
    \mathbf P\equiv\mathbf V^{-1}=
    \begin{pmatrix}
        \dfrac{1}{2(1+\kappa)} & \dfrac{1}{2(1+\kappa)} & \dfrac{\kappa\hn4}{(1+\kappa)(\hn3+\hn4)} & \dfrac{\kappa\hn3}{(1+\kappa)(\hn3+\hn4)}\\[10pt]
        \dfrac{1}{2(1+\kappa)} & \dfrac{1}{2(1+\kappa)} & -\dfrac{\hn4}{(1+\kappa)(\hn3+\hn4)} & -\dfrac{\hn3}{(1+\kappa)(\hn3+\hn4)}\\[10pt]
        1 & -1 & 0 & 0\\[4pt]
        0 & 0 & \dfrac{1}{\hn3+\hn4} & -\dfrac{1}{\hn3+\hn4}
    \end{pmatrix}
    \label{S:eq:Pmatrix}
\end{equation}
gives each mode as a combination of the four arms,
\begin{equation}
\begin{aligned}
    X&=\frac{1}{1+\kappa}\left[\frac{\hR_1+\hR_2}{2}+\kappa\,\frac{\hn4\hR_3+\hn3\hR_4}{\hn3+\hn4}\right],
    &\qquad Z_L&=\hR_1-\hR_2, \\[4pt]
    Y&=\frac{1}{1+\kappa}\left[\frac{\hR_1+\hR_2}{2}-\frac{\hn4\hR_3+\hn3\hR_4}{\hn3+\hn4}\right],
    &\qquad Z_T&=\frac{\hR_3-\hR_4}{\hn3+\hn4}.
\end{aligned}
\label{S:eq:modes_explicit}
\end{equation}

\sisub{Boundary conditions and the amplitude \texorpdfstring{$\kappa$}{kappa}}
\label{S:sec:bc_alpha}
Conditions (B1)--(B4) are eight scalar statements: (B1) fixes three independent differences, (B3) and (B4) two each, and (B2) one. Each is transferred to the modes by substitution into \eqref{S:eq:modes_explicit}, taking the modes in turn. In the fractional coordinate the two loop arms share the factor $1/N_L$, so the tension balance in (B3) reads $\partial_u\hR_1(1)+\partial_u\hR_2(1)=0$, and (B4) reads $\partial_u\hR_3(1)=\partial_u\hR_4(1)=0$.

\emph{$Z_L$, from (B1) and (B3).} Since $Z_L=\hR_1-\hR_2$, the junction continuity $\hR_1(0)=\hR_2(0)$ and the loop closure $\hR_1(1)=\hR_2(1)$ give
\begin{equation}
    Z_L(0)=0, \qquad Z_L(1)=0 .
    \label{S:eq:bc_ZL}
\end{equation}

\emph{$Z_T$, from (B1) and (B4).} Since $Z_T=(\hR_3-\hR_4)/(\hn3+\hn4)$, the junction continuity $\hR_3(0)=\hR_4(0)$ and the vanishing of both tail slopes at $u=1$ give
\begin{equation}
    Z_T(0)=0, \qquad \partial_uZ_T(1)=0 .
    \label{S:eq:bc_ZT}
\end{equation}

\emph{$Y$, from (B1), (B3) and (B4).} $Y$ is the difference between a loop-side and a tail-side average, each with weights summing to unity, so it vanishes at the junction, where (B1) makes the two averages equal. At $u=1$ the loop pair enters only through $\partial_u\hR_1(1)+\partial_u\hR_2(1)$, zero by the tension balance, and the tail pair through $\partial_u\hR_3(1)$ and $\partial_u\hR_4(1)$, zero by (B4). Hence
\begin{equation}
    Y(0)=0, \qquad \partial_uY(1)=0 .
    \label{S:eq:bc_Y}
\end{equation}

\emph{$X$, from (B3), (B4) and (B2).} $X$ combines the same two averages and inherits the far-end condition from the same two statements,
\begin{equation}
    \partial_uX(1)=0 .
    \label{S:eq:bc_X1}
\end{equation}

In the fractional coordinate, the statement (B2) reads $\sum_a\hn a^{-1}\partial_u\hR_a(0)=0$; inserting \eqref{S:eq:Rdef}, the $Z_L$ and $Z_T$ contributions cancel, $\hn L^{-1}(\tfrac12-\tfrac12)=0$ and $\hn3^{-1}\hn3-\hn4^{-1}\hn4=0$, leaving
\begin{equation}
    \partial_uX(0)\left[\frac{2}{\hn L}+\frac{1}{\hn3}+\frac1{\hn4}\right]
    +\partial_uY(0)\left[\frac{2\kappa}{\hn L}-\frac{1}{\hn3}-\frac1{\hn4}\right]=0 .
    \label{S:eq:B2modes}
\end{equation}
For independent boundary conditions on all modes, both terms must vanish separately. $Y$ is already fixed by \eqref{S:eq:bc_Y}, so no further condition can be imposed on $\partial_uY(0)$; its multiplier must therefore vanish, which uniquely determines $\kappa$. The bracket multiplying $\partial_uX(0)$ is strictly positive, so $\partial_uX(0)$ itself must vanish, giving the remaining boundary condition. Finally,
\begin{equation}
    \partial_uX(0)=0,
    \qquad
    \kappa=\frac{\hn L}{2}\left(\frac{1}{\hn3}+\frac1{\hn4}\right)
    \label{S:eq:kappa}
\end{equation}

\sinote{The drift matrix \texorpdfstring{$\A$}{A}}

\sisub{Mode equations by similarity transformation}
\label{S:sec:theta}
The transformation \eqref{S:eq:Vmatrix}--\eqref{S:eq:Pmatrix} has constant entries: $\mathbf V$ and $\mathbf P$ are built from $\kappa,\hn3,\hn4$ and depend on neither $u$ nor $\hat t$. As a result
\begin{equation}
    \partial_{\hat t}\Ps=\mathbf P\,\partial_{\hat t}\hR,
    \qquad
    \partial_u^2\hR=\mathbf V\,\partial_u^2\Ps .
    \label{S:eq:commute}
\end{equation}
Left-multiplying the arm system \eqref{S:eq:arm_matrix} by $\mathbf P$ gives
$\mathbf P\partial_{\hat t}\hR=\pi^{-2}\mathbf{PD}\,\partial_u^2\hR+\mathbf P\hf$, whose left side is already $\partial_{\hat t}\Ps$ and whose right side still refers to the arms. Eliminating them with the second identity of \eqref{S:eq:commute} closes the system in the modes alone:
\begin{equation}
    \partial_{\hat t}\Ps=\frac{1}{\pi^2}\,\bm\Theta\,\partial_u^2\Ps+\mathbf P\hf,
    \qquad
    \bm\Theta\equiv\mathbf P\,\mathbf D\,\mathbf V=\mathbf V^{-1}\mathbf D\,\mathbf V\
    \label{S:eq:mode_system}
\end{equation}
$\bm\Theta$ is the arm stiffness matrix carried into the mode basis, and $\mathbf P\hf$ is the arm noise carried into the same basis by the same matrix. With \eqref{S:eq:Vmatrix}, \eqref{S:eq:Pmatrix} and \eqref{S:eq:arm_matrix}, in the mode order $(X,Y,Z_L,Z_T)$,
\begin{equation}
    \setlength{\arraycolsep}{3pt}
    \bm\Theta\!=\!
    \begin{pmatrix}
        \dfrac{\hn L^{-2}+\kappa \big(\hn3^{-2}-\hn3^{-1}\hn4^{-1}+\hn4^{-2}\big)}{1+\kappa} & \dfrac{\kappa\big(\hn L^{-2}-\big(\hn3^{-2}-\hn3^{-1}\hn4^{-1}+\hn4^{-2}\big)\big)}{1+\kappa} & 0 & \dfrac{\kappa(\hn4-\hn3)}{(1+\kappa)\hn3\hn4}\\[12pt]
        \dfrac{\hn L^{-2}-\big(\hn3^{-2}-\hn3^{-1}\hn4^{-1}+\hn4^{-2}\big)}{1+\kappa} & \dfrac{\kappa\hn L^{-2}+\big(\hn3^{-2}-\hn3^{-1}\hn4^{-1}+\hn4^{-2}\big)}{1+\kappa} & 0 & \dfrac{\hn3-\hn4}{(1+\kappa)\hn3\hn4}\\[12pt]
        0 & 0 & \hn L^{-2} & 0\\[6pt]
        \dfrac{\hn4-\hn3}{(\hn3\hn4)^2} & \dfrac{\hn3-\hn4}{(\hn3\hn4)^2} & 0 & \dfrac{1}{\hn3\hn4}
    \end{pmatrix}\!.
    \label{S:eq:Theta_matrix}
\end{equation}
This shows that the matrix becomes completely diagonal when all arm lengths are equal (i.e. $\hn L=\hn 4=\hn 3$. $Z_T$ couplings are proportional to $\hn4-\hn3$ and vanish for tails of equal length. The $Z_L$ row and column vanish off the diagonal for every topology.

\sisub{Orthonormal spectral basis}
\label{S:sec:basis}
The operator $\partial_u^2$ under the conditions \eqref{S:eq:bc_ZL}--\eqref{S:eq:bc_X1} and \eqref{S:eq:kappa} has three distinct eigenbases, taken in orthonormal form:
\begin{equation}
    \phi^c_p(u)\equiv\sqrt2\cos(p\pi u), \qquad
    \phi^s_p(u)\equiv\sqrt2\sin(p\pi u), \qquad
    \phi^h_p(u)\equiv\sqrt2\sin\!\big((p-\tfrac12)\pi u\big),
    \label{S:eq:onb}
\end{equation}
normalized so that
\begin{equation}
    \int_0^1\phi^c_p\phi^c_q\dd=\int_0^1\phi^s_p\phi^s_q\dd=\int_0^1\phi^h_p\phi^h_q\dd=\delta_{pq}
    \label{S:eq:orthonormal}
\end{equation}
exactly. $X$ satisfies $\partial_uX(0)=\partial_uX(1)=0$ and is expanded in $\{\phi^c_p\}$; $Z_L$ satisfies $Z_L(0)=Z_L(1)=0$ and is expanded in $\{\phi^s_p\}$; $Y$ and $Z_T$ satisfy $Y(0)=0$, $\partial_uY(1)=0$, and $Z_T(0)=0$, $\partial_uZ_T(1)=0$, respectively and are expanded in $\{\phi^h_p\}$. Truncated at $p_{\max}$ terms,
\begin{equation}
\begin{aligned}
    X(u,\hat t)&=X_0(\hat t)+\sum_{p=1}^{p_{\max}}X_p(\hat t)\,\phi^c_p(u), &\qquad
    Z_L(u,\hat t)&=\sum_{p=1}^{p_{\max}}Z_{L,p}(\hat t)\,\phi^s_p(u), \\
    Y(u,\hat t)&=\sum_{p=1}^{p_{\max}}Y_p(\hat t)\,\phi^h_p(u), &\qquad
    Z_T(u,\hat t)&=\sum_{p=1}^{p_{\max}}Z_{T,p}(\hat t)\,\phi^h_p(u).
\end{aligned}
\label{S:eq:expand}
\end{equation}
The $p=0$ member of the cosine family is the constant $\phi^c_0\equiv1$. Writing $\phi^{(i)}_p$ for the basis function of mode $i$ and $\lambda^{(i)}_p$ for its eigenvalue,
\begin{equation}
    \partial_u^2\,\phi^{(i)}_p=-\pi^2\lambda^{(i)}_p\,\phi^{(i)}_p,
    \qquad
    \lambda^{(i)}_p=
    \begin{cases}
        p^2, & i=X,\;Z_L,\\[2pt]
        \big(p-\tfrac12\big)^2, & i=Y,\;Z_T,
    \end{cases}
    \label{S:eq:eigenvalues}
\end{equation}
collected into diagonal $p_{\max}\times p_{\max}$ matrices
\begin{equation}
    \bm\lambda^{(i)}\equiv\mathrm{diag}\big(\lambda^{(i)}_1,\dots,\lambda^{(i)}_{p_{\max}}\big),
    \qquad
    \bm\lambda^{c}=\bm\lambda^{s}=\mathrm{diag}\big(p^2\big),
    \qquad
    \bm\lambda^{h}=\mathrm{diag}\big((p-\tfrac12)^2\big).
    \label{S:eq:lammat}
\end{equation}
Because $X$ lives on an integer basis while $Y$ and $Z_T$ live on a half-integer basis, the two families are not mutually orthogonal; their overlap is
\begin{align}
    \Ihc(m,n)&\equiv\int_0^1\phi^h_m(u)\,\phi^c_n(u)\dd = 2\int_0^1\sin\!\big((m-\tfrac12)\pi u\big)\cos(n\pi u)\dd \notag\\
    &=\frac{4(2m-1)}{\pi\,(2m+2n-1)(2m-2n-1)}, \qquad \Ihc(m,n)\neq\Ihc(n,m),
    \label{S:eq:Ihc}
\end{align}
by the product-to-sum identity with $\cos\!\big((m-\tfrac12\pm n)\pi\big)=0$. The first index labels the half-integer function and the second the integer one, so $\int_0^1\phi^c_n\phi^h_m\dd=\Ihc(m,n)$ as well. As a $p_{\max}\times p_{\max}$ matrix the overlap
\begin{equation}
    \bm\Omega_{ij}\equiv\left[\int_0^1 \phi^{(i)}_q\phi^{(j)}_p\dd\right]_{qp}
    =\begin{cases}
        \mathbf I, & \text{$i$ and $j$ share a basis family},\\[2pt]
        \mathbf I_{hc}^{T}, & \phi^{(i)}=\phi^c,\;\phi^{(j)}=\phi^h,\\[2pt]
        \mathbf I_{hc}, & \phi^{(i)}=\phi^h,\;\phi^{(j)}=\phi^c,
    \end{cases}
    \qquad [\mathbf I_{hc}]_{qp}\equiv\Ihc(q,p),
    \label{S:eq:Omegamat}
\end{equation}
takes only three values. The pairs of $Z_L$ with $X$, $Y$ or $Z_T$ do not arise, because the corresponding entries of $\bm\Theta$ and $\bm\Gamma$ vanish, Eqs.~\eqref{S:eq:Theta_matrix} and \eqref{S:eq:Gamma_matrix}.

\sisub{Projection}
\label{S:sec:projection}
Expanding each component of $\Ps$ in its own basis replaces the four functions of $u$ by a $4\times p_{\max}$ array of coefficients $\Psi_{i,p}$, one row per mode. The state vector is that array read out row by row,
\begin{equation}
    \mathcal M=\big(X_1,\dots,X_{p_{\max}},\;Y_1,\dots,Y_{p_{\max}},\;Z_{L,1},\dots,Z_{L,p_{\max}},\;Z_{T,1},\dots,Z_{T,p_{\max}}\big)^{\!T},
    \label{S:eq:state}
\end{equation}
so the index pair $(i,p)$ of $\Psi_{i,p}$ maps onto the block structure of a $4p_{\max}\times4p_{\max}$ matrix: block $(i,j)$ couples row $i$ of the array to row $j$.

Taking the $i$-th row of \eqref{S:eq:mode_system}, multiplying by the test function $\phi^{(i)}_q$ of that same mode and integrating over $u\in[0,1]$, orthonormality \eqref{S:eq:orthonormal} reduces the left side to $\dot\Psi_{i,q}$ with no compensating factor. On the right, expanding $\Psi_j$ in its own basis and using \eqref{S:eq:eigenvalues},
\begin{equation}
    \frac{1}{\pi^2}\int_0^1\!\Theta_{ij}\,\partial_u^2\Psi_j\,\phi^{(i)}_q\dd
    =-\Theta_{ij}\sum_p\lambda^{(j)}_p\,\Psi_{j,p}\int_0^1 \phi^{(i)}_q\phi^{(j)}_p\dd ,
\end{equation}
the factors of $\pi^2$ cancelling. Summing over $j$ and writing $\tilde f_{i,q}\equiv\int_0^1\big[\mathbf P\hf\big]_i\,\phi^{(i)}_q\dd$,
\begin{equation}
    \dot\Psi_{i,q}=-\sum_j\sum_p \Theta_{ij}\,\lambda^{(j)}_p\,\big[\bm\Omega_{ij}\big]_{qp}\,\Psi_{j,p}\;+\;\tilde f_{i,q} .
    \label{S:eq:projected}
\end{equation}
Because the test functions are orthonormal no rescaling of $\tilde f$ is required: the noise entering \eqref{S:eq:projected} is exactly the projection of $\mathbf P\hf$. The scalar $\Theta_{ij}$ passes through the projection unopened, so \eqref{S:eq:projected} holds for any constant $4\times4$ matrix.

\sisub{Assembled matrix}
In the ordering \eqref{S:eq:state}, Eq.~\eqref{S:eq:projected} reads $\dot{\mathcal M}=-\A\,\mathcal M+\tilde{\mathbf f}$, each block being a product of three objects already defined:
\begin{equation}
    \mathbf A_{ij}=\Theta_{ij}\;\bm\Omega_{ij}\;\bm\lambda^{(j)}
    \qquad\text{i.e.}\qquad
    \big[\mathbf A_{ij}\big]_{qp}=\Theta_{ij}\,\lambda^{(j)}_p\,\big[\bm\Omega_{ij}\big]_{qp}\
    \label{S:eq:Ablock}
\end{equation}
Applying this to all sixteen pairs and removing the six blocks whose $\Theta_{ij}$ vanish by \eqref{S:eq:Theta_matrix},
\begin{equation}
    \A =
    \begin{pmatrix}
        \Theta_{XX}\,\bm\lambda^{c} & \Theta_{XY}\,\mathbf I_{hc}^{T}\bm\lambda^{h} & \mathbf 0 & \Theta_{XZ_T}\,\mathbf I_{hc}^{T}\bm\lambda^{h} \\[4pt]
        \Theta_{YX}\,\mathbf I_{hc}\bm\lambda^{c} & \Theta_{YY}\,\bm\lambda^{h} & \mathbf 0 & \Theta_{YZ_T}\,\bm\lambda^{h} \\[4pt]
        \mathbf 0 & \mathbf 0 & \Theta_{Z_LZ_L}\,\bm\lambda^{s} & \mathbf 0 \\[4pt]
        \Theta_{Z_TX}\,\mathbf I_{hc}\bm\lambda^{c} & \Theta_{Z_TY}\,\bm\lambda^{h} & \mathbf 0 & \Theta_{Z_TZ_T}\,\bm\lambda^{h}
    \end{pmatrix}.
    \label{S:eq:Amatrix}
\end{equation}
The diagonal blocks and the $Y$--$Z_T$ pair are diagonal matrices; the four blocks in the $X$ row and $X$ column are dense, and are the only dense blocks, all non-diagonal entries in those blocks arise due to interaction between the integer basis of $X$ and the half-integer basis of $Y,Z_T$. $\A$ is not symmetric.

\sinote{The noise covariance matrix \texorpdfstring{$\F$}{F}}
\label{S:sec:F}

\sisub{The amplitude matrix}
The mode noise is $\mathbf P\hf$ from \eqref{S:eq:mode_system}, the rows of $\mathbf P$ being its coefficients. Projecting onto the basis, $\tilde f_{i,q}=\int_0^1\big[\mathbf P\hf\big]_i\,\phi^{(i)}_q\dd$, and using \eqref{S:eq:fdt_dimless},
\begin{align}
    \big\langle \tilde f_{i,q}(\hat t)\,\tilde f_{j,p}(\hat t')\big\rangle
    &=\sum_{a,a'}P_{ia}P_{ja'}\int_0^1\!\!\int_0^1
      \big\langle \hf_a(u,\hat t)\cdot\hf_{a'}(u',\hat t')\big\rangle\,
      \phi^{(i)}_q(u)\,\phi^{(j)}_p(u')\dd\,du' \notag\\
    &=\sum_{a}P_{ia}P_{ja}\,g_a\,\delta(\hat t-\hat t')
      \int_0^1 \phi^{(i)}_q(u)\,\phi^{(j)}_p(u)\dd ,
    \label{S:eq:Fderiv}
\end{align}
the double arm sum collapsing to a single index because distinct arms are uncorrelated and the double $u$-integral collapsing on $\delta(u-u')$. The amplitude and the geometric factor separate, and the amplitude is a matrix product:
\begin{equation}
    \big\langle \tilde f_{i,q}(\hat t)\,\tilde f_{j,p}(\hat t')\big\rangle
    = \Gamma_{ij}\,\big[\bm\Omega_{ij}\big]_{qp}\,\delta(\hat t-\hat t'),
    \qquad
    \bm\Gamma\equiv\mathbf P\,\mathbf g\,\mathbf P^{T}\
    \label{S:eq:Gamma}
\end{equation}
with $\mathbf g\equiv\mathrm{diag}\big(2/(\pi^2\hn a)\big)$ the arm noise strengths of \eqref{S:eq:fdt_dimless}. The same $\mathbf P$ that produced $\bm\Theta=\mathbf{PDV}$ produces $\bm\Gamma$, and the same overlap \eqref{S:eq:Omegamat} serves both. $\bm\Gamma$ is symmetric, as a covariance must be, and has the same pattern of zeros as \eqref{S:eq:Theta_matrix},
\begin{equation}
    \setlength{\arraycolsep}{3pt}
    \bm\Gamma\!=\!\frac{2}{\pi^2}\!
    \begin{pmatrix}
        \dfrac{\dfrac{1}{2\hn L}+\kappa^2\dfrac{\hn3^2-\hn3\hn4+\hn4^2}{\hn3\hn4(\hn3+\hn4)}}{(1+\kappa)^2} & \dfrac{\dfrac{1}{2\hn L}-\kappa \dfrac{\hn3^2-\hn3\hn4+\hn4^2}{\hn3\hn4(\hn3+\hn4)}}{(1+\kappa)^2} & 0 & \dfrac{\kappa(\hn4-\hn3)}{(1+\kappa)\hn3\hn4(\hn3+\hn4)}\\[14pt]
        \dfrac{\dfrac{1}{2\hn L}-\kappa \dfrac{\hn3^2-\hn3\hn4+\hn4^2}{\hn3\hn4(\hn3+\hn4)}}{(1+\kappa)^2} & \dfrac{\dfrac{1}{2\hn L}+\dfrac{\hn3^2-\hn3\hn4+\hn4^2}{\hn3\hn4(\hn3+\hn4)}}{(1+\kappa)^2} & 0 & \dfrac{\hn3-\hn4}{(1+\kappa)\hn3\hn4(\hn3+\hn4)}\\[14pt]
        0 & 0 & \dfrac{2}{\hn L} & 0\\[10pt]
        \dfrac{\kappa(\hn4-\hn3)}{(1+\kappa)\hn3\hn4(\hn3+\hn4)} & \dfrac{\hn3-\hn4}{(1+\kappa)\hn3\hn4(\hn3+\hn4)} & 0 & \dfrac{1}{\hn3\hn4(\hn3+\hn4)}
    \end{pmatrix}\!,
    \label{S:eq:Gamma_matrix}
\end{equation}
for the same reason: the two loop arms have equal length and hence equal noise strength, and only the $Z_L$ row of $\mathbf P$ is antisymmetric in them.

\sisub{Assembled matrix}
With $\langle\tilde{\mathbf f}(\hat t)\tilde{\mathbf f}^{T}(\hat t')\rangle=\F\,\delta(\hat t-\hat t')$,
\begin{equation}
    \mathbf F_{ij}=\Gamma_{ij}\;\bm\Omega_{ij}\
    \label{S:eq:Fblock}
\end{equation}
the same rule as \eqref{S:eq:Ablock} with $\bm\Gamma$ in place of $\bm\Theta$ and no eigenvalue factor, the noise being white. Written out,
\begin{equation}
    \F =
    \begin{pmatrix}
        \Gamma_{XX}\,\mathbf I & \Gamma_{XY}\,\mathbf I_{hc}^{T} & \mathbf 0 & \Gamma_{XZ_T}\,\mathbf I_{hc}^{T} \\[4pt]
        \Gamma_{XY}\,\mathbf I_{hc} & \Gamma_{YY}\,\mathbf I & \mathbf 0 & \Gamma_{YZ_T}\,\mathbf I \\[4pt]
        \mathbf 0 & \mathbf 0 & \Gamma_{Z_LZ_L}\,\mathbf I & \mathbf 0 \\[4pt]
        \Gamma_{XZ_T}\,\mathbf I_{hc} & \Gamma_{YZ_T}\,\mathbf I & \mathbf 0 & \Gamma_{Z_TZ_T}\,\mathbf I
    \end{pmatrix}\,.
    \label{S:eq:Fmatrix}
\end{equation}
The symmetry of $\F$ is manifest: $\mathbf I$ is symmetric and the off-diagonal partners $\mathbf I_{hc}^{T}$ and $\mathbf I_{hc}$ are transposes carrying the same scalar, since $\Gamma_{ij}=\Gamma_{ji}$. Both conditions fail for $\A$.

\sinote{The state covariance \texorpdfstring{$\mathbf C$}{C}}
\label{S:sec:covariance}

The projected system
\begin{equation}
    \dot{\mathcal M}=-\A\,\mathcal M+\tilde{\mathbf f},
    \qquad
    \big\langle\tilde{\mathbf f}(\hat t)\tilde{\mathbf f}^T(\hat t')\big\rangle=\F\,\delta(\hat t-\hat t')
    \label{S:eq:OU}
\end{equation}
is a multivariate Ornstein--Uhlenbeck process. Every eigenvalue of $\A$ has positive real part, so it relaxes to a unique stationary state, in which
\begin{equation}
    \mathcal M(\hat t)=\int_{-\infty}^{\hat t}e^{-\A(\hat t-\hat t_1)}\,\tilde{\mathbf f}(\hat t_1)\,d\hat t_1 .
    \label{S:eq:OUsol}
\end{equation}
Two covariances are required, the equilibrium matrix and the time-dependent one,
\begin{equation}
    \mathbf C(0)\equiv\big\langle\mathcal M(0)\mathcal M^T(0)\big\rangle,
    \qquad
    \mathbf C(\hat t)\equiv\big\langle\mathcal M(\hat t)\mathcal M^T(0)\big\rangle,\quad \hat t\ge0.
    \label{S:eq:Cdefs}
\end{equation}
Because the process has run since $\hat t=-\infty$, the instant $\hat t=0$ is no different from any other: $\mathbf C(0)$ is the equilibrium covariance at every time, and the argument of $\mathbf C$ is a time lag. Both covariances follow from \eqref{S:eq:OUsol}, which at $\hat t=0$ reads
\begin{equation}
    \mathcal M(0)=\int_{-\infty}^{0}e^{\A\hat t_1}\,\tilde{\mathbf f}(\hat t_1)\,d\hat t_1 .
    \label{S:eq:M0}
\end{equation}

\sisub{Equilibrium covariance}
The outer product of \eqref{S:eq:M0} with itself is a double integral over the noise. The exponentials are not random, so averaging acts on the noise alone and leaves $\F\,\delta(\hat t_1-\hat t_2)$ by \eqref{S:eq:OU}, which collapses the double integral onto $\hat t_2=\hat t_1$. Substituting $\tau=-\hat t_1$ then gives
\begin{align}
    \mathbf C(0)
    &=\int_{-\infty}^{0}\!\!\int_{-\infty}^{0} e^{\A\hat t_1}\,\big\langle\tilde{\mathbf f}(\hat t_1)\tilde{\mathbf f}^{T}(\hat t_2)\big\rangle\,e^{\A^{T}\hat t_2}\,d\hat t_1\,d\hat t_2
    \nonumber\\
    &=\int_{-\infty}^{0} e^{\A\hat t_1}\,\F\,e^{\A^{T}\hat t_1}\,d\hat t_1
    =\int_{0}^{\infty} e^{-\A\tau}\,\F\,e^{-\A^{T}\tau}\,d\tau .
    \label{S:eq:C0int}
\end{align}
Two properties follow by inspection. The integrand is $\F$ conjugated by $e^{-\A\tau}$ and its transpose, hence symmetric at every $\tau$, so $\mathbf C(0)=\mathbf C(0)^{T}$; and $\mathbf x^{T}e^{-\A\tau}\F e^{-\A^{T}\tau}\mathbf x=\mathbf y^{T}\F\mathbf y\ge0$ with $\mathbf y=e^{-\A^{T}\tau}\mathbf x$, so $\mathbf C(0)\succeq0$.

Equation~\eqref{S:eq:C0int} is a complete expression for $\mathbf C(0)$, but it is a quadrature over matrix exponentials. One step reduces it to an algebraic equation. Write the integrand as $\mathbf J(\tau)\equiv e^{-\A\tau}\F e^{-\A^{T}\tau}$ and differentiate; each exponential commutes with its own generator and contributes a factor on the side it stands,
\begin{equation}
    \frac{d\mathbf J}{d\tau}
    =-\A\,e^{-\A\tau}\F e^{-\A^{T}\tau}-e^{-\A\tau}\F e^{-\A^{T}\tau}\,\A^{T}
    =-\A\,\mathbf J(\tau)-\mathbf J(\tau)\,\A^{T}.
    \label{S:eq:dK}
\end{equation}
Integrating \eqref{S:eq:dK} over $\tau\in[0,\infty)$, the right-hand side is linear in $\mathbf J$, so the constant matrices come outside the integral and what remains of it is $\mathbf C(0)$ by \eqref{S:eq:C0int}:
\begin{equation}
    \mathbf J(\infty)-\mathbf J(0)
    =-\A\int_{0}^{\infty}\mathbf J\,d\tau-\left(\int_{0}^{\infty}\mathbf J\,d\tau\right)\A^{T}
    =-\Big[\A\,\mathbf C(0)+\mathbf C(0)\,\A^{T}\Big].
    \label{S:eq:Kint}
\end{equation}
On the left $\mathbf J(0)=\F$, and $\mathbf J(\infty)=\mathbf 0$ because every eigenvalue of $\A$ has positive real part, so $e^{-\A\tau}\to\mathbf 0$. Hence
\begin{equation}
    \A\,\mathbf C(0)+\mathbf C(0)\,\A^{T} = \F\
    \label{S:eq:lyapunov}
\end{equation}
a continuous-time Lyapunov equation.

\sisub{Closed-form equilibrium covariance}
\label{S:sec:C0closed}
\emph{One matrix for drift and noise.} Collect the arm lengths into $\mathbf N\equiv\mathrm{diag}(\hn L,\hn L,\hn3,\hn4)$. The arm stiffness and noise matrices of \eqref{S:eq:arm_matrix} and \eqref{S:eq:fdt_dimless} are then $\mathbf D=\mathbf N^{-2}$ and $\mathbf g=(2/\pi^2)\,\mathbf N^{-1}$. Inserting $\mathbf I=\mathbf P^{T}\mathbf V^{T}$, which holds because $\mathbf P=\mathbf V^{-1}$, between the two factors of $\mathbf N^{-1}$ in $\bm\Theta=\mathbf P\mathbf N^{-1}\mathbf N^{-1}\mathbf V$,
\begin{equation}
    \bm\Theta=\big(\mathbf P\mathbf N^{-1}\mathbf P^{T}\big)\big(\mathbf V^{T}\mathbf N^{-1}\mathbf V\big)
    =\frac{\pi^2}{2}\,\bm\Gamma\,\mathbf k,
    \qquad
    \mathbf k\equiv\mathbf V^{T}\mathbf N^{-1}\mathbf V,
    \label{S:eq:Theta_Gamma_k}
\end{equation}
where $\bm\Gamma=\mathbf P\mathbf g\mathbf P^{T}=(2/\pi^2)\,\mathbf P\mathbf N^{-1}\mathbf P^{T}$ is the noise amplitude matrix \eqref{S:eq:Gamma}. The matrix $\mathbf k$ is the elastic stiffness of the arms written in the mode coordinates: by \eqref{S:eq:Rdef}, $\sum_a\hn a^{-1}(\partial_u\hR_a)^2=\partial_u\Ps^{T}\,\mathbf k\,\partial_u\Ps$.

\emph{The stiffness is diagonal.} With $\mathbf V$ from \eqref{S:eq:Vmatrix}, in the mode order $(X,Y,Z_L,Z_T)$, the only off-diagonal entry of $\mathbf k$ that does not vanish identically is $k_{XY}=2\kappa/\hn L-1/\hn3-1/\hn4$, the bracket of \eqref{S:eq:B2modes}, and it is zero for the amplitude ratio \eqref{S:eq:kappa}. Hence
\begin{equation}
    \setlength{\arraycolsep}{10pt}
    \mathbf k=\mathrm{diag}\big(k_X,k_Y,k_{Z_L},k_{Z_T}\big)=
    \begin{pmatrix}
        \dfrac{2}{\hn L}+\dfrac{1}{\hn3}+\dfrac{1}{\hn4} & 0 & 0 & 0\\[4pt]
        0 & \dfrac{2\kappa^2}{\hn L}+\dfrac{1}{\hn3}+\dfrac{1}{\hn4} & 0 & 0\\[4pt]
        0 & 0 & \dfrac{1}{2\hn L} & 0\\[4pt]
        0 & 0 & 0 & \hn3+\hn4
    \end{pmatrix}.
\label{S:eq:k_entries}
\end{equation}

\emph{Factorization of the drift.} By \eqref{S:eq:Theta_Gamma_k}, $\Theta_{ij}=(\pi^2/2)\Gamma_{ij}k_j$, and every block of \eqref{S:eq:Ablock} becomes a block of \eqref{S:eq:Fblock} times a diagonal matrix,
\begin{equation}
\begin{gathered}
    \mathbf A_{ij}=\Theta_{ij}\bm\Omega_{ij}\bm\lambda^{(j)}=\frac{\pi^2}{2}\,\Gamma_{ij}\bm\Omega_{ij}\,k_j\bm\lambda^{(j)},
    \qquad\text{i.e.}\qquad
    \A=\frac{\pi^2}{2}\,\F\,\K,\\[4pt]
    \K\equiv\mathrm{blockdiag}\big(k_X\bm\lambda^{c},\,k_Y\bm\lambda^{h},\,k_{Z_L}\bm\lambda^{s},\,k_{Z_T}\bm\lambda^{h}\big).
\end{gathered}
    \label{S:eq:AFK}
\end{equation}

\emph{Solution of the Lyapunov equation.} Since $\F$ is symmetric and $\K$ is diagonal, \eqref{S:eq:AFK} gives
\begin{equation}
    \A\,\K^{-1}=\frac{\pi^2}{2}\,\F=\K^{-1}\A^{T}:
    \label{S:eq:AKsym}
\end{equation}
$\A$ is not symmetric, but $\A\K^{-1}$ is. Writing the right-hand side of \eqref{S:eq:lyapunov} as the sum of its two equal halves, $\F=\pi^{-2}\big(\A\K^{-1}+\K^{-1}\A^{T}\big)$, the Lyapunov equation becomes
\begin{equation}
    \A\,\mathbf C(0)+\mathbf C(0)\,\A^{T}=\A\,\frac{\K^{-1}}{\pi^2}+\frac{\K^{-1}}{\pi^2}\,\A^{T}.
    \label{S:eq:lyap_AK}
\end{equation}
The right-hand side has the same form as the left, with $\mathbf C(0)$ replaced by $\K^{-1}/\pi^2$. 
Therefore
\begin{equation}
    \mathbf C(0)=\frac{\K^{-1}}{\pi^2},
    \qquad
    \big\langle\Psi_{i,p}\Psi_{j,q}\big\rangle=\frac{\delta_{ij}\delta_{pq}}{\pi^2\,k_i\,\lambda^{(i)}_p},
    \label{S:eq:C0closed}
\end{equation}
suggesting that at equal times the mode amplitudes are independent of each other.

\sisub{Time-dependent covariance}
\label{S:sec:Ct_sym}
Differentiating $\mathbf C(\hat t)$ of \eqref{S:eq:Cdefs} in its first argument and using \eqref{S:eq:OU},
\begin{equation}
    \frac{d\mathbf C}{d\hat t}=\big\langle\dot{\mathcal M}(\hat t)\,\mathcal M^T(0)\big\rangle
    =-\A\,\mathbf C(\hat t)+\big\langle\tilde{\mathbf f}(\hat t)\,\mathcal M^T(0)\big\rangle .
    \label{S:eq:Ct_ode_raw}
\end{equation}
Here the second factor is held at time $0$, and for $\hat t>0$ the noise term vanishes: by \eqref{S:eq:OUsol} the state $\mathcal M(0)$ is built from noise at times before $0$, and the noise is delta-correlated, so $\tilde{\mathbf f}(\hat t)$ is uncorrelated with it. The equation that remains is homogeneous,
\begin{equation}
    \frac{d\mathbf C}{d\hat t}+\A\,\mathbf C(\hat t)=\mathbf 0,\qquad \hat t>0 .
    \label{S:eq:Ct_ode}
\end{equation}
Multiplying by the integrating factor $e^{\A\hat t}$ collects the two terms into one derivative,
\begin{equation}
    \frac{d}{d\hat t}\Big[e^{\A\hat t}\,\mathbf C(\hat t)\Big]
    =e^{\A\hat t}\left[\frac{d\mathbf C}{d\hat t}+\A\,\mathbf C(\hat t)\right]=\mathbf 0 ,
    \label{S:eq:intfactor}
\end{equation}
so $e^{\A\hat t}\mathbf C(\hat t)$ is constant on $\hat t\ge0$ and equal to its value $\mathbf C(0)$ at $\hat t=0$. Hence
\begin{equation}
    \mathbf C(\hat t)=e^{-\A\hat t}\,\mathbf C(0), \qquad \hat t\ge0,\
    \label{S:eq:Ct}
\end{equation}
with $\mathbf C(0)$ supplied by \eqref{S:eq:C0closed}. The time-dependent covariance decays under the same operator that governs the mean. The restriction to $\hat t\ge0$ is where causality entered: for $\hat t<0$ it is the noise at time $0$ that is uncorrelated with the earlier state, and differentiating in the second argument instead gives $\mathbf C(0)e^{\A^T\hat t}$.

Although $\A$ is not symmetric, $\mathbf C(\hat t)$ is: by \eqref{S:eq:AKsym}, $e^{-\A\hat t}\K^{-1}=\K^{-1}e^{-\A^{T}\hat t}$, so with \eqref{S:eq:C0closed} $\mathbf C(\hat t)^T=\mathbf C(\hat t)$, the time-reversal symmetry of an equilibrium process. Unlike $\mathbf C(0)$, it is not diagonal.

\sinote{Radius of gyration}
\label{S:sec:rg}

\sisub{Centre of mass in mode variables}
The centre of mass is the contour-weighted average over the four arms,
\begin{equation}
    \hR_{cm}(\hat t)=\sum_{a=1}^4\hn a\int_0^1\hR_a(u,\hat t)\dd ,
    \label{S:eq:Rcm_def}
\end{equation}
the weights summing to unity by \eqref{S:eq:fracs}. By \eqref{S:eq:Rdef} and \eqref{S:eq:expand}, the position on arm $a$ is
\begin{equation}
    \hR_a(u,\hat t)=X_0(\hat t)+\sum_{i}\sum_{p=1}^{p_{\max}}V_{ai}\,\Psi_{i,p}(\hat t)\,\phi^{(i)}_p(u),
    \label{S:eq:armcompact}
\end{equation}
with $i$ running over $X,Y,Z_L,Z_T$ and $X_0$ the rigid translation. Inserting \eqref{S:eq:armcompact} into \eqref{S:eq:Rcm_def},
\begin{equation}
    \hR_{cm}(\hat t)=X_0(\hat t)+\sum_{i}\sum_{p=1}^{p_{\max}}\Big(\sum_{b=1}^4\hn b V_{bi}\int_0^1\phi^{(i)}_p\dd\Big)\Psi_{i,p}(\hat t).
    \label{S:eq:Rcm_modes}
\end{equation}
The bracket vanishes for $X$, whose cosines integrate to zero for $p\ge1$, and for $Z_L$, since $\sum_b\hn bV_{bZ_L}=0$. For $Y$ and $Z_T$ it is the product of the weighted column sums $\sum_b\hn bV_{bY}=2\hn L\kappa-\hn3-\hn4$ and $\sum_b\hn bV_{bZ_T}=\hn3^2-\hn4^2$ with
\begin{equation}
    \rho_p\equiv\int_0^1\phi^h_p(u)\dd=\frac{\sqrt2}{\big(p-\tfrac12\big)\pi}.
    \label{S:eq:rho}
\end{equation}
The rigid-translation amplitude $X_0$ is therefore not the centre of mass: the butterfly and tail modes carry a net contour-weighted displacement, vanishing only when $2\hn L\kappa=\hn3+\hn4$ and $\hn3=\hn4$ respectively.

\sisub{Definition and closed form}
\label{S:sec:Rg_closed}
The radius of gyration is
\begin{equation}
    \big\langle \hat R_g^2\big\rangle=\sum_{a=1}^4\hn a\int_0^1\big\langle\big(\hR_a(u)-\hR_{cm}\big)^2\big\rangle\dd .
    \label{S:eq:Rg_def}
\end{equation}
By \eqref{S:eq:armcompact} and \eqref{S:eq:Rcm_modes}, the deviation from the centre of mass is a fixed linear combination of the state vector,
\begin{equation}
    \hR_a(u)-\hR_{cm}=\mathbf B_a(u)\,\mathcal M,
    \label{S:eq:Bproj}
\end{equation}
with the non-fluctuating row vector $\mathbf B_a(u)$ of length $4p_{\max}$, whose entry multiplying $\Psi_{i,p}$ is
\begin{equation}
    \big[\mathbf B_a(u)\big]_{i,p}=V_{ai}\,\phi^{(i)}_p(u)-\sum_{b=1}^4\hn b\,V_{bi}\int_0^1\phi^{(i)}_p\dd .
    \label{S:eq:Bvec}
\end{equation}
The first term is the mode field on arm $a$; the second is its contour average over the chain, which implements the centre-of-mass subtraction \eqref{S:eq:Rcm_modes}. At the junction every half-integer and integer sine vanishes and every cosine equals $\sqrt2$, so all four $\mathbf B_a(0)$ coincide.

\emph{Mode sum.} By \eqref{S:eq:Bproj}, $\hR_a(u)-\hR_{cm}=\sum_{i,p}[\mathbf B_a(u)]_{i,p}\Psi_{i,p}$, with $[\mathbf B_a(u)]_{i,p}$ from \eqref{S:eq:Bvec}. By \eqref{S:eq:C0closed} the mode amplitudes are uncorrelated at equal times, $\langle\Psi_{i,p}\Psi_{j,q}\rangle=\delta_{ij}\delta_{pq}\,C_{i,p}$ with $C_{i,p}=1/(\pi^2k_i\lambda^{(i)}_p)$, so squaring and averaging in \eqref{S:eq:Rg_def} leaves only the diagonal terms,
\begin{equation}
    \big\langle\hat R_g^2\big\rangle=\sum_i\sum_{p}C_{i,p}\,G_{i,p},
    \qquad
    G_{i,p}\equiv\sum_{a=1}^4\hn a\int_0^1\big[\mathbf B_a(u)\big]_{i,p}^2\dd .
    \label{S:eq:Rg_sum}
\end{equation}

\emph{Geometric factors.} Squaring \eqref{S:eq:Bvec}, integrating with the orthonormality \eqref{S:eq:orthonormal} of the basis functions and using $\sum_a\hn a=1$,
\begin{equation}
    G_{i,p}=\sum_{a=1}^4\hn a V_{ai}^2-\Big(\sum_{a=1}^4\hn a V_{ai}\Big)^2\Big(\int_0^1\phi^{(i)}_p\dd\Big)^2 .
    \label{S:eq:Gip}
\end{equation}
With $\mathbf V$ from \eqref{S:eq:Vmatrix} and $\rho_p$ from \eqref{S:eq:rho}, $G_{X,p}=1$, $G_{Z_L,p}=\tfrac12\hn L$, $G_{Y,p}=2\hn L\kappa^2+\hn3+\hn4-\big(2\hn L\kappa-\hn3-\hn4\big)^2\rho_p^2$ and $G_{Z_T,p}=\hn3^3+\hn4^3-\big(\hn3^2-\hn4^2\big)^2\rho_p^2$. Products of different entries of $\mathbf B_a$, such as the $X$--$Y$ overlaps involving $\Ihc$, do not contribute, because the corresponding covariances vanish. The remaining sums over $p$ in \eqref{S:eq:Rg_sum} are elementary, $\sum_{p\ge1}p^{-2}=\pi^2/6$, $\sum_{p\ge1}(p-\tfrac12)^{-2}=\pi^2/2$ and $\sum_{p\ge1}\rho_p^2(p-\tfrac12)^{-2}=\pi^2/3$, and were carried out analytically for $p_{\max}\to\infty$. The result reduces to $1/6$ for an unlooped chain ($\hn L\to0$) and to $1/12$ for a ring ($\hn3,\hn4\to0$).

\sinote{Mean-square displacement}
\label{S:sec:msd}

\sisub{Centre-of-mass motion}
Differentiating \eqref{S:eq:Rcm_def} and inserting \eqref{S:eq:rouse_dimless},
\begin{equation}
    \dot{\hR}_{cm}=\sum_a\hn a\int_0^1\!\left[\frac{\partial_u^2\hR_a}{\pi^2\hn a^2}+\hf_a\right]\!\dd
    =\sum_a\frac{1}{\pi^2\hn a}\Big[\partial_u\hR_a(1)-\partial_u\hR_a(0)\Big]
    +\sum_a\hn a\int_0^1\hf_a\dd .
\end{equation}
The boundary terms vanish identically: at $u=1$, (B3) gives $\partial_u\hR_1(1)+\partial_u\hR_2(1)=0$, the two loop arms carrying the common factor $1/\hn L$, and (B4) gives $\partial_u\hR_3(1)=\partial_u\hR_4(1)=0$; at $u=0$, (B2) is $\sum_a\hn a^{-1}\partial_u\hR_a(0)=0$. Hence
\begin{equation}
    \dot{\hR}_{cm}(\hat t)=\sum_a\hn a\int_0^1\hf_a(u,\hat t)\dd ,
    \label{S:eq:Rcm_dot}
\end{equation}
exactly: the centre of mass performs free Brownian motion driven by noise alone.

Its driving noise is uncorrelated with every internal mode noise. Using $\mathbf P\hf$ and \eqref{S:eq:fdt_dimless},
\begin{equation}
    \big\langle\dot{\hR}_{cm}(\hat t)\,\tilde f_{i,q}(\hat t')\big\rangle
    =\left(\sum_a\hn a P_{ia}\,g_a\right)\delta(\hat t-\hat t')\int_0^1\phi^{(i)}_q(u)\dd .
    \label{S:eq:cm_orth}
\end{equation}
For $i=X$ the integral vanishes, $\int_0^1\phi^c_q\dd=0$ for $q\ge1$. For $i=Y$ the coefficient sum vanishes, the loop and tail contributions being $2\pi^{-2}/(1+\kappa)$ and $-2\pi^{-2}/(1+\kappa)$; likewise for $i=Z_L$ and $i=Z_T$, whose coefficients are equal and opposite on the two arms concerned. Therefore $\hR_{cm}(\hat t)$ and $\mathcal M(\hat t)$ are independent processes and the displacement of any locus decomposes without a cross term,
\begin{equation}
    \big\langle\Delta\hR_a^2(u,\hat t)\big\rangle
    =\big\langle\Delta\hR_{cm}^2(\hat t)\big\rangle
    +\big\langle\big(\Delta\hat{\mathbf r}_a(u,\hat t)\big)^2\big\rangle,
    \qquad
    \hat{\mathbf r}_a\equiv\hR_a-\hR_{cm}.
    \label{S:eq:MSD_split}
\end{equation}
From \eqref{S:eq:Rcm_dot} and \eqref{S:eq:fdt_dimless}, using $\sum_a\hn a=1$,
\begin{equation}
    \big\langle\dot{\hR}_{cm}(\hat t)\cdot\dot{\hR}_{cm}(\hat t')\big\rangle
    =\sum_a\hn a^2\,\frac{2}{\pi^2\hn a}\,\delta(\hat t-\hat t')
    =\frac{2}{\pi^2}\,\delta(\hat t-\hat t'),
\end{equation}
a pure number, with no topological dependence. Integrating and matching to the three-dimensional free-diffusion form $\langle\Delta\hR^2\rangle=6\hat D_{cm}\hat t$,
\begin{equation}
    \big\langle\Delta\hR_{cm}^2(\hat t)\big\rangle=\frac{2}{\pi^2}\hat t
    \;\Longrightarrow\;
    \hat D_{cm}=\frac{1}{3\pi^2},
    \label{S:eq:Dcm}
\end{equation}
the dimensionless form of the Rouse result $D_{cm}=k_BT/(N_{tot}\zeta)$ under the scales \eqref{S:eq:scales}.

\sisub{Internal displacement}
\label{S:sec:internal}
By \eqref{S:eq:Bproj}, the internal displacement of the locus $u$ on arm $a$ is a fixed linear combination of the state vector, $\hat{\mathbf r}_a(u,\hat t)=\mathbf B_a(u)\,\mathcal M(\hat t)$, with the row vector $\mathbf B_a(u)$ of \eqref{S:eq:Bvec}.

Since $\mathbf B_a(u)$ is a single row and does not fluctuate it passes outside every average. Writing $\mathbf B$ for $\mathbf B_a(u)$ and using stationarity for the equilibrium term,
\begin{equation}
    \big\langle\hat{\mathbf r}(\hat t)\cdot\hat{\mathbf r}(\hat t)\big\rangle=\big\langle\hat{\mathbf r}(0)\cdot\hat{\mathbf r}(0)\big\rangle=\mathbf B\,\mathbf C(0)\,\mathbf B^T,
    \qquad
    \big\langle\hat{\mathbf r}(\hat t)\cdot\hat{\mathbf r}(0)\big\rangle=\mathbf B\,\mathbf C(\hat t)\,\mathbf B^T ,
\end{equation}
and expanding the mean-square increment gives
\begin{equation}
    \big\langle\big(\Delta\hat{\mathbf r}_a(u,\hat t)\big)^2\big\rangle
    = 2\,\mathbf B_a(u)\big[\mathbf C(0)-\mathbf C(\hat t)\big]\mathbf B_a(u)^T.
    \label{S:eq:MSD_internal}
\end{equation}
Since $\mathbf C(\hat t)$ is symmetric (Sec.~\ref{S:sec:Ct_sym}), the two time orderings of the cross term agree.

\sisub{Complete forward model}
Combining \eqref{S:eq:MSD_split}, \eqref{S:eq:Dcm}, \eqref{S:eq:MSD_internal} and \eqref{S:eq:Ct},
\begin{equation}
    \big\langle\Delta\hR_a^2(u,\hat t)\big\rangle
    = \frac{2}{\pi^2}\hat t
    + 2\,\mathbf B_a(u)\Big[\mathbf I-e^{-\A\hat t}\Big]\mathbf C(0)\,\mathbf B_a(u)^T,
    \label{S:eq:MSD_full}
\end{equation}
an exact function of $\hat t$, of the locus $(a,u)$, and of the two anchor fractions $(s_a,s_b)$ alone, through $\hn3,\hn L,\hn4$ and hence $\kappa$, $\bm\Theta$, $\bm\Gamma$ and $\mathbf C(0)$, with no adjustable parameter.

The quantity reported is the local scaling exponent, the logarithmic slope of \eqref{S:eq:MSD_full}. Its only $\hat t$-dependence sits in the exponential, which differentiates to $-\A e^{-\A\hat t}$, so
\begin{equation}
    \frac{d}{d\hat t}\big\langle\Delta\hR_a^2(u,\hat t)\big\rangle
    =\frac{2}{\pi^2}+2\,\mathbf B_a(u)\,\A\,e^{-\A\hat t}\,\mathbf C(0)\,\mathbf B_a(u)^T ,
    \label{S:eq:MSDdot}
\end{equation}
and the exponent follows as a ratio of two closed forms,
\begin{equation}
    \alpha(\hat t)\equiv
    \frac{d\ln\big\langle\Delta\hR_a^2\big\rangle}{d\ln\hat t}
    =\frac{\hat t}{\big\langle\Delta\hR_a^2\big\rangle}\,
     \frac{d\big\langle\Delta\hR_a^2\big\rangle}{d\hat t}.
    \label{S:eq:exponent}
\end{equation}
Taken this way $\alpha(\hat t)$ carries no differencing error and no dependence on how the time axis is sampled.

\sipart{Methods}
\label{S:sec:implementation}

\sinote{Implementation}
\label{S:sec:implementation_details}
For given anchor fractions $(s_a,s_b)$ the fractional lengths were obtained from \eqref{S:eq:fracs} and the amplitude $\kappa$ from \eqref{S:eq:kappa}. The matrix $\mathbf V$ was formed from \eqref{S:eq:Vmatrix} and inverted numerically for $\mathbf P$; $\mathbf D$ and $\mathbf g$ were formed from \eqref{S:eq:arm_matrix} and \eqref{S:eq:fdt_dimless}. The mode-space matrices were evaluated as the products $\bm\Theta=\mathbf{PDV}$ and $\bm\Gamma=\mathbf{PgP}^{T}$ rather than from closed-form entries. The equilibrium covariance was evaluated from the closed form \eqref{S:eq:C0closed}, with no Lyapunov solve, and $\mathbf C(\hat t)=e^{-\A\hat t}\mathbf C(0)$ of \eqref{S:eq:Ct} from one eigendecomposition of the drift matrix per topology, $\A=\mathbf U\,\mathrm{diag}(\sigma_m)\,\mathbf U^{-1}$, reused for every time requested. Its eigenvalues, the relaxation rates $\sigma_m$, were real and positive in every case, to round-off. For a locus with projection row $\mathbf B$ the MSD \eqref{S:eq:MSD_full} then reads
\begin{equation}
    \big\langle\Delta\hR^2(\hat t)\big\rangle=\frac{2}{\pi^2}\,\hat t+2\sum_m w_m\big(1-e^{-\sigma_m\hat t}\big),
    \qquad
    w_m=\big(\mathbf B\mathbf U\big)_m\big(\mathbf U^{-1}\mathbf C(0)\mathbf B^T\big)_m ,
    \label{S:eq:CtS}
\end{equation}
and its time derivative, used for $\alpha$, follows term by term. Observables followed from \eqref{S:eq:Rg_sum} and from \eqref{S:eq:Bvec} with \eqref{S:eq:MSD_full}.

The modes of all four arms share one basis in the rescaled coordinate $u$, so the mode of index $p$ resolves a contour length $\hn a/p$ on arm $a$ and relaxes there on the time scale $(\hn a/p)^2$. The truncation is therefore coarsest on the longest arm, $\hn{\max}=\max(\hn L,\hn3,\hn4)$, and the relative truncation error of the MSD scales as $(\hn{\max}/p_{\max})/\sqrt{\hat t}$. We require $\hat t\ge10^{3}(\hn{\max}/p_{\max})^2$ at every time reported, and for each topology therefore chose
\begin{equation}
    p_{\max}=\Big\lceil \hn{\max}\sqrt{10^{3}/\hat t_{\min}}\,\Big\rceil ,
    \label{S:eq:pmax}
\end{equation}
with $\hat t_{\min}$ the shortest time reported: $10^{-5}$ for Fig.~3(a)--(c), $1.5\times10^{-5}$ for the dip analysis (Sec.~\ref{S:sec:dip}), and $10^{-4}$ for the inference library (Sec.~\ref{S:sec:inference}), with at least 40 modes. Equilibrium quantities involve no time scale and were computed with $p_{\max}=1000$ for contacts and shape anisotropy, which resolves contour separations far below the smallest one considered. $\langle\hat R_g^2\rangle$ was evaluated from \eqref{S:eq:Rg_sum} with the sums over $p$ carried out analytically.

\sinote{Equilibrium configurations}
\label{S:sec:configs}
The stationary distribution of the state vector $\mathcal M$ is Gaussian with zero mean and covariance $\mathbf C(0)$, Eq.~\eqref{S:eq:C0closed}. Its three Cartesian components are independent and identically distributed, each with covariance $\mathbf C(0)/3$. A configuration was generated by drawing each component as $\mathcal M_\gamma=\mathbf L\,\bm\xi_\gamma$, $\gamma=x,y,z$, where $\mathbf L$ is the Cholesky factor of $\mathbf C(0)/3$, $\mathbf L\mathbf L^T=\mathbf C(0)/3$, and the $\bm\xi_\gamma$ are vectors of independent standard normal variables. The arm positions follow from the mode amplitudes through Eqs.~\eqref{S:eq:Rdef} and \eqref{S:eq:armcompact}, and their positions relative to the centre of mass, Eq.~\eqref{S:eq:Rcm_modes}, through the projection \eqref{S:eq:Bproj}, $\hat{\mathbf r}_a(u)=\mathbf B_a(u)\,\mathcal M$. Loci were placed at uniformly spaced contour fractions $s$ and assigned to arms by Eq.~\eqref{S:eq:s_to_u}; the two anchors map to the same point, the junction. For the unlooped chain, the amplitudes of the Rouse modes $\sqrt2\cos(p\pi s)$ were drawn independently, with variance $1/(3\pi^2p^2)$ per Cartesian component.

For each configuration, the gyration tensor $\mathbf T=n_s^{-1}\sum_{i}\hat{\mathbf r}_i\hat{\mathbf r}_i^{T}$ was formed from the $n_s$ loci, and its eigenvalues $\Lambda_1,\Lambda_2,\Lambda_3$ give the squared radius of gyration, $\Lambda_1+\Lambda_2+\Lambda_3$, and the relative shape anisotropy $\chi^2$ defined in the main text. Each value of $\langle\chi^2\rangle$ in Fig.~2(d) of the main text averages $10^4$ configurations with loci spaced by $0.01$.

The configurations of Fig.~2(a) of the main text were selected as follows. For each topology, $300$ configurations were drawn with loci spaced by $0.001$, and the one minimizing $z_{R}^2+z_{\chi}^2$ was retained, where $z_{R}$ and $z_{\chi}$ are the deviations of its $R_g^2$ and $\chi^2$ from the means over the $300$ configurations, in units of the corresponding standard deviations. The retained configuration was projected onto the plane spanned by the eigenvectors of its gyration tensor with the two largest eigenvalues. The curve drawn is a coarse-grained version of the same configuration, obtained by setting to zero all amplitudes with mode index $p>50$ in each of the four mode families before the projection \eqref{S:eq:Bproj}.

\sinote{Dip analysis}
\label{S:sec:dip}
The onset and minimum times of Fig.~3(d)--(f) of the main text were computed for centred loops, $n_c=0.5$, with $\hn L=0.05,0.10,\dots,0.25$. For each loop, tagged loci were placed at contour distance $d$ from the junction on arm~1 (loop side) at ten logarithmically spaced values from $d=0.01$ to the loop centre, $d=\hn L$, and on arm~3 (tail side) at ten logarithmically spaced values from $d=0.01$ to the free end, $d=\hn3$. The exponent $\alpha(\hat t)$ was evaluated from Eq.~\eqref{S:eq:exponent} at $400$ logarithmically spaced times from $\hat t\approx 1.5\times10^{-5}$ to $\hat t=10$.

The onset time $t_{dev}$ is the first time at which $\alpha<1/2-\delta\alpha$, and $t_{min}$ is the time at which $\alpha$ is minimal within the time window. The margin $\delta\alpha=0.01$ separates a genuine dip from the slow approach of $\alpha$ to $1/2$ at early times.

A least-squares fit of $\ln t_{dev}=\ln c+2\ln d$ is the geometric mean of $t_{dev}/d^2$; over all loci of Fig.~3(e) of the main text, on both sides of the junction and for all loop sizes, it gives $c\approx3.4$. The same fit to $t_{min}$ for loci on the loop close to the anchor, $d/\hn L<0.1$, gives $t_{min}\approx54\,d^2$. The value $t_{min}\approx3.6\,\hn L^2$ near the loop centre is read from the plateau of the collapsed data in Fig.~3(f) of the main text.

\sinote{Inference of the anchors}
\label{S:sec:inference}
The library contains the MSD, Eq.~\eqref{S:eq:MSD_full}, of the loci $s=0,0.01,\dots,1$, assigned to arms by Eq.~\eqref{S:eq:s_to_u}, at the six times of the main text, for all $N=4095$ candidate topologies; a locus at an anchor is the junction. The slider positions $x=0.20,0.21,\dots,0.70$ and the loop lengths $2\hn L=0.02,0.04,\dots,0.20$ keep every locus and every true topology on this grid. The measured values $g_k^*$ are the library values of the true topology, without added noise, so the true topology is itself a candidate, with $\mathcal{D}^2=0$.

\end{document}